\documentclass[12pt,preprint]{aastex}

\shorttitle{HH9779}
\shortauthors{Austin, Robertson, Tycner}

\begin{document}

\title{Late-Type Near-Contact Eclipsing Binary [HH97] FS Aur-79}

\author{S. J. Austin}
\affil{Department of Physics and Astronomy, University of Central Arkansas,
    Conway, AR 72035}
\email{saustin@uca.edu}

\author{J. W. Robertson}
\affil{Department of Physical Sciences, Arkansas Tech University, 
Russellville AR 72801-2222}
\email{jeff.robertson@atu.edu}

\author{C. Tycner\altaffilmark{1,2}}
\affil{US Naval Observatory, Flagstaff Station, Flagstaff, AZ 86001-8521}
\email{tycner@sextans.lowell.edu}

\author{T. Campbell}
\affil{Whispering Pines Observatory, Harrison AR 72601}
\email{jmontecamp@yahoo.com}

\and

\author{R. K. Honeycutt}
\affil{Department of Astronomy, Indiana University, Bloomington, IN 47405}
\email{honey@astro.indiana.edu}

\altaffiltext{1}{Michelson Postdoctoral Fellow.}
\altaffiltext{2}{NVI, Inc. 7257 Hanover Parkway, Suite D, Greenbelt, MD 20770}

\begin{abstract}

The secondary photometric standard star \#79 for the FS Aur field (Henden \& Honeycutt 1997) 
designated as [HH97] FS Aur-79 (GSC 1874 399) is a short period (0.2508 days) eclipsing 
binary whose light curve is a combination of the $\beta$ Lyr and BY Dra type variables.  
High signal-to-noise multi-color photometry were obtained using the USNO 1-m telescope.  
These light curves show asymmetry at quadrature phases (O'Connell effect), 
which can be modeled with the presence of star spots.  
A low resolution spectrum obtained with the 3.5-m WIYN telescope at orbital phase 0.76 is 
consistent with a spectral type of dK7e and dM3e.  
A radial velocity curve for the primary star was constructed using twenty-four high resolution spectra
from the 9.2 m HET.
Spectra show H$\alpha$ and H$\beta$  in emission confirming chromospheric activity and possibly 
the presence of circumstellar material.  
Binary star models that simultaneously fit the U, B, V, R  and RV curves are those with 
a primary star of mass 0.59$\pm$0.02 M$_\sun$, temperature 4100$\pm$25 K, mean radius of 0.67 R$_\sun$, 
just filling its Roche lobe and a secondary star of mass 0.31$\pm$0.09 M$_\sun$, temperature 3425$\pm$25 K, 
mean radius of 0.48 R$_\sun$, just within its Roche lobe.
An inclination angle of  83\degr\ $\pm$2\degr\ with a center of mass separation of 1.62 R$_\sun$ is 
also derived.  
Star spots, expected for a rotation period of less than a day,  had to be included in the modeling to fit 
the O'Connell effect.

\end{abstract}


\section{Introduction}

Henden and Honeycutt (1997) designated this star as the 79th standard star in the 
field of cataclysmic variable star FS Aur and this star is also cataloged GSC 1874 399
 in the Guide Star Catalogue and mis-referenced as [HH95] FS-79 by SIMBAD\footnote{
The SIMBAD database is operated at the CDS in Strasbourg France}, but should be [HH97] FS-79.
For this paper we will refer to this star as FS Aur-79.  
The discovery that this star was actually an eclipsing binary was described by 
Robertson et al. (2004).  
They found that the orbital period was approximately
0.25 days and that the temperatures of the two stars were not equal, indicating
that the two stars were not in contact.  
Their low signal-to-noise BVR photometry from 2003-2004 enabled them to determine
preliminary system parameters such as mass ratio, inclination angle, effective
temperatures, and filling factors.  
This indicated a near contact binary system with late-type components.  

Good, accurate, fundamental parameters for late-type main sequence stars are based 
on only a handful of double-lined spectroscopic, totally eclipsing binaries with 
K and M components.  
Therefore there are considerable gaps in our interpolations and knowledge.  
This is also true when determining effective temperature for cooler M dwarfs as the complexity of molecules present in their atmospheres increases with later spectral types.
Near contact and contact systems with these very late components are extremely rare 
and are therefore interesting for what they can tell us about binary stellar formation 
and evolution.

To better constrain the system parameters we obtained high signal-to-noise ratio (SNR) UBVR
photometry, and spectra to determine the radial velocity curves.
In Sec. 2 we describe how the photometry and spectroscopy were obtained and reduced.
In Sec. 3 we present the photometric and spectroscopicl analysis.
Sec. 4 is a discussion of our results, and Sec. 5 is a summary of our conclusions.

\section{Observations and Reductions}

\subsection{Photometry}

\subsubsection{Center for Backyard Astrophysics Photometry}

After the initial discovery of this eclipsing variable star (Robertson et al. 2004), 
additional light curves were collected by Tut Campbell (CBA-Arkansas) on 15 nights 
during 2003-November through 2004-February.  
The interest of the Center for Backyard Astrophysics (CBA) in the cataclysmic variable FS Aur,
 very near our newly discovered variable, 
also helped spark a professional-amateur observational campaign by some members of the 
CBA to collect differential photometry for  obtaining times of eclipse minima.  
Data were gathered by Tut Campbell (CBA-Arkansas), Tom Krajci (2006) (CBA-New Mexico), 
Pierre de Ponthiere, (CBA-Lesve), and Jerry Foote (CBA-Utah) on 45 nights from 
2005-October to 2006-January.  Details of the observatories and their 
telescope/detector configurations can be found at the CBA\footnote{\tt http://cba.phys.columbia.edu/}.  
These light curves were utilized to obtain measurements of the times of minima for both primary and 
secondary eclipses that can be found in Table 1.

\subsubsection{USNO Photometry}

Photometric observations obtained at the U. S. Naval
Observatory~(USNO), Flagstaff Station, were acquired using the 1-m
Ritchey-Chr\'{e}tien telescope.  
The telescope was equipped with a Tektronix 2048$\times$2048 
thinned back-illuminated CCD and Johnson $UBV$ filters.  
The camera has a scale of 0.68 arcsec pixel$^{-1}$,
which results in a field of view of 23.2 by 23.2 arcmin.

The field centered on FS Aur was observed on three consecutive nights,
2005 Dec 4, 5, and 6~(UT), in $B$, $V$, and $U$ filters, respectively.
Resulting light curves are shown in Figures 1-3.
Each set of nightly observations was preceded with a set of 20 to 50
bias frames and 10 dome flats in all three filters.  
A total of 110, 176, and 73 individual exposures were obtained in the $B$, $V$, and
$U$ filters, respectively.  
Each night the field was observed during an 8~hr window, which corresponded to hour angles 
between $-4$ to $+4$~hr.  
Although each night was devoted to a single filter, at least
one set of exposures in all three filters was also acquired each
night.  
The typical seeing during the USNO observing run was between 2 and 3 arcsec.

\subsubsection{R Band Photometry}

Time-series differential R-band photometry of FS Aur-79 was obtained on 2006 Jan 2 at the Whispering Pines Observatory in Harrison Arkansas using a 0.4-m f4.2 Newtonian telescope with a Johnson R filter and SBIG ST-6 CCD camera.
The differential R-band light curve is shown in Figure 4.
The R data is very valuable as the extra color information helps reduce the uniqueness problem and limit parameter space in modeling binary light curves, especially those with star spots.

\subsubsection{Photometric Reductions}

All of the CCD images containing photometry data were processed using 
{\tt Cmuniwin}\footnote{\tt http://integral.sci.muni.cz/cmunipack/index.html}.  
This PC-based software with a graphical user interface was converted from Linux C programs 
originally developed by Filip Hroch (1998).  
In short, the images are 1) converted to FITS 
if necessary, 2) flat-fielded and dark subtracted if desired, 3) processed to find stellar 
targets and photometrically measured utilizing the algorithms of DAOPHOT (Stetson 1987), 4) 
target lists are pattern matched to identify stars in each image via the algorithm of Groth, 
(1986), and 5) variable, comparison and check stars selected to generate differential photometry 
and light curves.

FS Aur-79 lies close to the CV FS Aur, (Robertson, et al. 2004), and the field has been observed to establish secondary standard stars (Henden \& Honeycutt 1997; Henden 2006).  
Therefore, there are comparison and check stars in the field with standard U, B, and V magnitudes 
but not R as of yet.  
The magnitude of the comparison star was added to the differential photometry values to yield the 
magnitudes displayed in Figures 1-3.

\subsection{Spectroscopy}

\subsubsection{WIYN}

Three spectra were obtained on 22 Feb 2006 UT with the HYDRA Multiple Object Spectrograph on the 
WIYN\footnote{The WIYN observatory is a joint facility of the University of Wisconsin-Madison, Indiana University, 
Yale University and the National Optical Astronomy Observatory} 3.5-m telescope at Kitt Peak National 
Observatory.  
Table 2 lists the HJD and orbital phase for each of the spectra obtained.
The 600 mm$^{-1}$ grating blazed at $13.9\degr$ was used in the first order to obtain a spectrum covering 
$\approx$ 535.0-825.0 nm pix$^{-1}$ with a resolution of $\approx$ 0.14 nm pix$^{-1}$.  
Exposure times for the object spectra were 600 seconds with wispy cirrus clouds present.  
Dome flat and HeArNe arc lamp images were obtained for calibrations to remove individual 
CCD pixel variations and to convert pixels to wavelengths.  
Reductions utilized standard IRAF\footnote{IRAF is distributed by the National Optical Astronomy Observatory, which is operated by the 
Association of Universities for Research in Astronomy, Inc. under cooperative agreement with the 
National Science Foundation} {\tt onedspec}/{\tt twodspec} procedures to 
1) create average flat field, and bias frames ({\tt imcombine}), 
2) subtract the bias from object and calibration images as well as flat field correct the images 
({\tt imarith}), 
3) identify, trace, sum, and extract 1-D apertures from the 2-D CCD images ({\tt apall}), 
4) determine and apply the wavelength calibration to object apertures from the HeArNe lamp 
apertures ({\tt identify}, {\tt refspec}), 
5) apply corrections to eliminate the motion of the observatory ({\tt rvcorr}, {\tt dopcor}), and 
6) correct the spectra to a flat continuum for relative comparison ({\tt continuum}).  
Although some unassigned fibers yield sky, the sky was not extracted or subtracted.  

Spectra of FS Aur-62 and FS Aur-63 were also obtained simultaneously through fibers as these are the
comparison stars used in the differential photometry.
The photometric calibrations of these stars (Henden \& Honeycutt 1997) 
shows that the $B-V$ colors of these stars are 0.412 and 0.450 respectively.
To relative flux calibrate the FS Aur-79 spectra we used the spectra of comparison stars with similar
$B-V$ colors:  LTT 377 ($B-V=$0.48), LTT1788 ($B-V=$0.47), and LTT2415 ($B-V=$0.40).
We then used spectra of FS Aur-62 and FS Aur-63 with the calibrations of those spectrosocopic
standards to obtain a relative flux calibration of the FS Aur-79 spectrum using IRAF tasks {\tt standard}, {\tt sensfunc}, and {\tt calibrate}.

The average spectrum of FS Aur-79 is shown in Figure 5 and corresponds to orbital phase $\approx0.766$.
The lines were identified using the line list for K and M stars from Kirkpatrick et al. (1991).

\subsubsection{HET HRS Spectroscopy}

Radial velocity measurements were needed to determine the individual stellar masses.
Given the approximately six hour orbital period, a visual magnitude of about 14, 
and the need to acquire high resolution spectra, a large aperture telescope was needed.
We therefore obtained 10 hours of priority-2 time during 2005 semester B (September through January)
on the High Resolution Spectrograph (HRS) at the 9.2 m Hobby-Eberly Telescope (HET) through NOAO.   
Due to the nature of the HET service observing scheduling only one object spectrum per
half night was possible.  
In order to get as complete sampling over an orbit as possible, we requested that
observations be done as randomly in time as possible over the semester, avoiding 
consecutive nights at the same time of night due to the approximatley six hour period
of the orbit.  
The HRS configuration yielding a resolving power of 15,000, wavelength coverage of 
approximately 4076\AA\ to 7838\AA, a fiber with an angular diameter on the sky of
3\arcsec, one sky fiber, no image slicer, no iodine gas cell, and binning of 2$\times$5 for
3.2 pixels per resolving elements were used.
Bias, flat, and Th-Ar comparison lamp frames were also obtained by the night observers for
reduction and calibration purposes.  
Spectral standards were also obtained, giving us the option to relative flux calibrate the spectra.
Table 3 lists the details of the 24 HET HRS exposures.

Standard IRAF tasks were used to process the frames into one-dimensional spectra normalized to the 
continuum.
The following reduction steps were used for both the red and blue channels:
{\tt zerocombine}, {\tt flatcombine}, {\tt ccdproc} to trim and apply the zero correction, 
{\tt cosmicrays}, {\tt apall} used to find and define the aperatures for each spectral order on the
red and blue chips, {\tt apscatter} to remove any scattered light, {\tt apflatten} to get
a flat field solution using the combined flat, and {\tt ccdproc} to apply the flattening
to the spectral frames, {\tt apall} once again to extract the spectral orders.
HET HRS orders 78-145 were extracted corresponding to spectral coverage from 4200\AA\ to 7900\AA.
When neccessary sky substraction was done for those nights near full moon when the sky 
background was significant and for those orders which have problematic sky lines 
(i.e. order 104, NaI-D doublet). 
Wavelength calibration was done for each spectral order of interest using {\tt ident},
{\tt reident}, {\tt refspec}, and {\tt dispcor}. 
The spectra where then converted to the heliocentric frame of reference using {\tt rvcor}
and {\tt dopcor}.  
The instrumental flux was then normalized to the continuum by using either {\tt contin}
if it could correctly find the continuum, or by doing the continuum fitting interactively
using the option under {\tt splot}.
Spectra from order 94 are shown in Figure 6 sorted by phase.
The main spectral feature in this order is the absorption line blend of
Ti I $\lambda$6497.689\AA, Ba I $\lambda$6498.759\AA, Fe I $\lambda$6498.950\AA,
Ca I $\lambda$6499.694\AA, Fe I $\lambda$6501.681\AA.
Figure 7 shows spectra from order 104 containing the NaI-D doublet sorted by phase.

\section{Analysis}

\subsection{Ephemeris}

Both primary and secondary eclipse times of minima were determined using the method of Kwee and
 van Woerden (1956) and are listed in Table 1.  
All of these timings were then used in a linear least-squares fit of the form $T=T_0+PN$ to refine 
the period and ephemeris, where $T_0$ is the heliocentric julian date of primary eclipse minimum (occultation of 
the more massive, hotter star) from Robertson et. al. (2004), $N$ is the 
orbital cycle since the epoch $T_0$, $P$ the orbital period, and $T$, predicted times of primary eclipses.  
This yielded a period of $0.250816(1)$ days.  
The differences between observed eclipses and their calculated timings (O-C) based on this new 
ephemeris $MIN\ I = HJD\ 2,452,963.74445 + 0.250816N$, are shown in Figure 8.
Given that (O-C) has been approximately zero over four years of photometry would indicate no
changes in the orbital period and therefore no indication of any significant mass transfer.

\subsection{Light Curves}

The U, B, V, and R phased light curves of FS Aur-79 are displayed in Figures 9-12.
The morphology of these light curves place FS Aur-79 into the the catagory of $\beta$ Lyrae eclipsing
binaries since the light curves vary continuously and the primary (phase=0.00) and secondary eclipses (phase=0.50) are of unequal depths.
Table 4 lists the UBVR magnitudes at primary eclipse, secondary eclipse, Max I (phase=0.25), and 
Max II (phase=0.75).
By comparing the magnitude of the secondary eclipse to the magnitudes during the quadratures we 
are able to find a lower limit to the magnitude differences between the primary and secondary star, since we do not know if the eclipses are total.
The primary star is at least 1.42 magnitudes brighter than the secondary in V,
and at least 1.33 magnitudes brighter than the secondary in B.

The light curves exhibit the O'Connell effect (unequal maxima) of BY Dra variables indicating star spots.
The amount of the O'Connell effect in each color can be quantified by the difference ($\Delta m$) between Max II and Max I , and are listed in Table 5.
Also shown in Table 5 are the $\Delta m$  U-B, B-V, and V-R colors.

Color curves were computed by first interpolating each light curve to a phase grid with phase increments of 0.0125, then subtracting the light curves of different colors, and therefore also different nights.
Figure 13 shows the U-B, B-V, and differential V-(R-comp) colors with respect to orbital phase.
Table 4 includes the U-B and B-V colors at primary eclipse, secondary eclipse, and Max I and Max II.
During secondary eclipse the secondary star is mostly hidden, and the observed color will be for the
primary star.
B-V during secondary eclipse is about 1.29, which would correspond to a main-sequence star with an effective temperature
of around 4100 K and spectral type of about K7.
During secondary eclipse U-B=0.860 which would correspond to a main-sequence star
with effective temperature of about 4700K.
This U-B excess is an indicator of chromospheric activity (which is also seen in our spectra) and therefore this U-B cannot be
used to determine an accurate photospheric temperature (Amado 2003).

The apparent V magnitude of the primary is  13.829, and the B-V color is 1.292.
Assuming the primary is a main-sequence star, then the distance modulus $V-M=5.82$ and the distance
is about 146 pc (similar distance to the Pleiades).
Given this distance, reddening and extinction is probably small enough to be ignored.

\subsection{Spectra}

\subsubsection{Spectral Classification}

The WIYN spectrum was obtained when the orbital phase of the system was about 0.76.  
Figure 14 shows our WIYN spectra, a main-sequence K7 spectrum (star HD237903), and 
a main-sequence M1 spectrum (star HD204445) from the 
Indo-U.S. Library of Coud\'{e} Feed Spectra\footnote{\tt http://www.noao.edu/cflib/}
(Valdes et al. 2004).
The K7 and M1 spectra were broadened with a rotational kernel calculated for a vsin(i) 
value of 135 km sec$^{-1}$ (approximate rotational speed of FS Aur-79) and a limb
darkening coefficient of 0.6 (see example for example Gray (2005) page 465).
The match between the WIYN spectra and the K7V spectrum is very close between
5200\AA\ and 7000\AA.
Between 7000\AA\ and 7500\AA\ the WIYN spectrum looks a bit more like an early dM spectrum.
But the spectrum between 7000\AA\ and 7500\AA\  is mostly a blend of a K7V spectrum and a  M type spectrum.
Given the uncertainties involved in the atmospheres of cool M-type stars and the lack of absolute flux calibration 
available, it is difficult at this time to accurately pinpoint the precise spectral type of the secondary from the WIYN spectrum.  
Binary modeling in section 3.3 indicate a secondary temperature consistent with an M4V star,
But, the mass and radius found by our binary modeling are more consistent with approximately 
an M2V or M3V star by comparing to the masses and radii determined for YY Gem and  CM Dra (Strassmeier et al. 1988).
Of all the potential spectroscopic chromospheric indicators that are accessible in our wavelength 
coverage like emission in hydrogen Balmer lines, CaII H,K emission cores, HeI 5876 absorption or 
emission core, NaI-D doublet emssion cores, etc., this star shows strong emission in H$\alpha$ 
along with corresponding emission in H$\beta$.
We therefore estimate the spectral types as dK7e for the primary star and dM3e for the secondary star.

\subsubsection{Velocities}

FS Aur-79 is a single lined spectroscopic binary with the lines from the primary star
most obvious, and the presence of the lines of the secondary detected by their influence
on the line profiles and TiO band strengths.
Given the luminosity difference (in the V band the secondary is at most 0.27 times as bright as the secondary), the short period, and  the short exposure times prohibited us from resolving the lines of the secondary.

The $\gamma$ velocity (radial velocity of the center of mass of the system relative 
to the Sun), and the radial velocities of the
primary can be measured directly.
We first determined the $\gamma$-velocity by choosing lines that we were confident were not blends.
For example, Ca I lines whose rest wavelengths are 6122.219 \AA\ and 6162.172 \AA.
Wavelengths of spectral features of interest were measured using the equivalent width function
in the IRAF task {\tt splot}.
A sine wave was fit to the wavelength versus orbital phase for these lines which determined the
wavelength of the line in the center of mass frame relative to the Sun, 6123.94$\pm$0.07 \AA\ and
6163.78$\pm$0.11 \AA\ respectively resulting in $\gamma$-velocities of +84.3$\pm$ 3.6 km sec$^{-1}$ and +78.3$\pm$5.4  km sec$^{-1}$.
The chromospheric H$\alpha$ line does not show any modulation of its wavelength as a function of orbital
phase, so it too was used to determine the $\gamma$-velocity.
Here a line with slope zero was fit to the wavelength versus orbital phase data resulting
in an observed wavelength of 6564.68$\pm$0.12 \AA\ and a $\gamma$-velocity of +85.2$\pm$5.5  km sec$^{-1}$.
We conclude that the $\gamma$-velocity of the system, based on the mean of the values above, 
is +82.6$\pm$5.5 km sec$^{-1}$.

We ran the IRAF cross-correlation task {\tt fxcor}  on HET HRS orders 94 thru 100, covering 
a wavelength range of 6070\AA\ to 6535 \AA, which contains several strong lines.  
We used a good signal-to-noise ratio spectrum near orbital phase 0.5 as the template.  
The resulting radial velocity curve had unacceptably large radial velocities errors.
The large errors are likely the result of the spectra changing with phase due to not only
doppler shifting, but also due to star spot groups, and to a lesser extent the changing relative contributions to the light from the two stars due to the eclipses.  
By measuring the wavelengths of over a dozen strong absorption lines using the equivalent width function in the IRAF task {\tt splot}, we were not as sensitive to these other effects.  
Figure 15 shows an example radial velocity curve from the absorption line of of the blend at 6500\AA\  in the center of mass frame of the system.
The errorbars in phase indicate the exposure time for each spectrum and the 
errorbars in velocity indicate the spectral resolution of the HRS spectra.
The maximum radial velocity (vsin(i)) of the primary is about 125$\pm$15 km sec$^{-1}$.
Also shown in Figure 15 are the radial velocity curves of both the primary and secondary stars from 
the best-fit binary model  discussed section 3.4 below.

\subsection{Binary Modeling}

\subsubsection{Optimization}

Light and radial velocity curves were fit simultaneously using the program {\tt Nightfall}\footnote{\tt http://www.hs.uni-hamburg.de/DE/Ins/Per/Wichmann/Nightfall.html}.
{\tt Nightfall} has built in optimization procedures (simplex and simulated annealing).
Temperature of the primary derived from the photometry and spectroscopy was set to 4100K.
The orbital period was set to 0.2508 days.
Model atmosphere, detailed reflection, and a square root limb darkening law options were enabled.
The temperature of the secondary was initally set to a value consistent with a early dM star.
We then had {\tt Nightfall} search parameter space allowing it to vary the temperature of the
secondary,  mass ratio, inclination angle, and filling factors for the primary and secondary.  
Initially only spotless models were computed.
{\tt Nightfall} converged on values of mass ratio, inclination angle, and filling factor listed in Table 6.
Nightfall computes a reduced chi-square  (${\chi_\nu}^2$) of the simultaneous fit of light and radial velocity curves for the best spotless model of approximately 54.

Spotless models could not reproduce the O'Connell effect and therefore various star-spot models were tried.
Star spots are to be expected for stars that have a rotation period of only 0.2508 days (Bopp \& Fekel 1977; Young et al. 1987).
The temperature of the secondary, and the longitude, size, and temperature factor relative to the photosphere (dimming factor) of the circular spot were allowed to vary.
The longitude convention used in {\tt Nightfall} has 0\degr\ on the secondary facing L1 and on the primary 0\degr\  facing away from L1.
Optimized single-equatorial-spot on the secondary model resulted in a ${\chi_\nu}^2$ of about 22.
The best single-spot on the secondary model parameters are also listed in Table 6.

Although this single spot model is an improvement over the spotless model, one spot is inadequate to account for the O'Connell effect seen.
In other words, the photospheric temperature inhomogeneity  is more complex than a single star spot.
In fact, what we are modeling as single spots are probably large spot groups.
The UBVR light curves were fit the best by a three spots model (Figures 9-12), one circular spot on the secondary and two 
circular spots on the primary  (Figure 16).
Again the optimization varied the temperature of the secondary, and the longitude, size, and dimming factor
for the spots and resulted in a fit with a ${\chi_\nu}^2$ of about 9.
Table 6 also shows the best-fit parameters for the three spot model obtained using {\tt Nightfall}.
The uncertainties listed in Table 6 represent the parameter ranges over which minimum ${\chi_\nu}^2$ 
remains essentially unchanged. 
We also compared the modeled and observed U-B, B-V, and V-(R-comp) color curves (Figure 13).
The B-V curve fits the observed values closely at all phases.
The U-B curve fits generally well, but the observed data shows features that are not reproduced by the model.
The V-(R-comp) also fits quit well, but the model seems to have systematic difference compared to the data.
The actual spots are probably spot groups which are irregular in shape,
and therefore the model cannot reproduce in detail the sunspot group shapes. 

The model parameters optimized by Nightfall were cross-checked by modeling with BinaryMaker 3.0 ({\tt BM3})
(Bradstreet \&\ Stellman 2002).
{\tt BM3} assumes the stars to be blackbodies, where as the stellar atmosphere option was used for the {\tt Nightfall} models. 
To achieve a fit similar to that obtained with {\tt Nightfall} (Figures 9-12) we had to change the inclination angle slightly and use a secondary temperature 50 K cooler when using {\tt BM3}, otherwise the same parameter values were used.
The equivalent best-ft three-spot parameters used in BinaryMaker are shown in Table 7.
As can be seen in Figures 9-12, the U and B light curves from {\tt BM3} do not fit as well as the {\tt Nightfall} equivalent, 
but the V and R light curves are closer to being identical.
We attribute these differences to the fact that model atmospheres were used for the {\tt Nightfall} modeling and
blackbodies for the {\tt BM3} model.

By simultaneously fitting the U, B, V, and R light curves we improved the likelihood of finding a unique
solution.  
One should interpret the star spots in the model simply as a way to introduce temperature differences
on the surfaces of the stars, since one cannot distinguish a cool spot from a hot spot on the opposite
side of the system.
By comparing the observed and modeled color curves (U-B, B-V, and V-R) in Figure 13, one can see
that the model does a good job of reproducing the observed temperature inhomogeneities.

\subsubsection{Spectral Synthesis}

Given the spectral types of the component stars and their observed and predicted radial velocities
we have synthesized the Na-D line profile for zero and quadrature orbital phases.
This provides a qualitative check on the orbital velocity of the secondary star whose spectral features
are not noticable from visual inspection of the HET HRS spectra.
Figure 17 shows the observed and synthesized profiles for order 100.  
A K7V spectrum and an M1V spectrum were normalized to the continuum over the wavelength range of 
the HRS HET order.  The M1V spectrum was then scaled by a factor of about 0.225
(relative luminosity of an M1V star in units of a K7V star's luminosity).
The composite spectrum was then normalized to the continuum.
The result mimics the line profiles of the sodium doublet, which indicates that the secondary star
spectrum is having an observable effect on this spectral feature, and that the predicted secondary
velocity is probably close to the actual velocity.

\section{Discussion}

FS Aur-79 represents a very near-contact system containing late-type (K-M dwarf) components. 
Near contact systems have been defined by Shaw (1990), in which the binaries have a 
period of less than 1 day, show strong tidal interactions, and have surfaces very near one 
another but not in contact.  
These stars are related to W UMa type systems and are probable 
precursors to such systems.  
W UMa systems have periods less than 1 day and their light curves show distortions due 
to their close proximity, similar to near-contact systems.  
There are no known W UMa type systems with an M star component, so this system represents 
a unique precursor to such a low-mass W UMa system.  

Unlike the W UMa systems, however, the near contact systems show unequal eclipse depths 
in their primary and secondary minima.
This can be explained with the fact that the stars have different temperatures and are
therefore not in contact and hence not in thermal equilibrium, indicative of $\beta$ Lyr systems.

There has been some speculation as to the actual existence of W UMa type systems which 
are currently not in the contact state due to thermal relaxations (Lucy 1976; Flannery 
1976; Wang 1994) or that the stars are somehow able to maintain temperature differences within a contact discontinuity  (Shu et al. 1976; Wang 1999).
These proposed systems have been referred to as the B-type (apart from W-type and 
A-type) subclass of W UMa systems. 
Indeed it has been shown that some of the observational candidates for the B-type W UMa 
systems can be the result of spurious effects produced when star spots are not included in light
curve models.
Maceroni \&\ Van 'T Veer (1990) showed the effect of modeling a near contact system with and without 
star spots.
Their results indicated that a spotless contact model could reproduce equally well the light curves of spotted non-contact systems. 

Star spots are notorious for introducing ambiguity in light curve modeling in 
chromospherically active binaries such as RS CVn and Algols as well as BY Dra rotational 
variables, especially when there is only one color band light curve.  
Multi-wavelength studies, coupled with suitable spectroscopy (i.e. radial velocities 
and phase correlated chromospheric activity indicators) can often disentangle multiple 
light curve models with the larger number of free parameters introduced by inclusion of star spots.

Close binaries with late-type components are very rare for two reasons, the first and 
most obvious is their intrinsic faintness and therefore observational bias in detecting 
and observing them in the first place.  
The other is due to their evolutionary history.  
Theories of protostellar collapse involving formation of close binary stars with 
separations ~1 A.U. or less are problematic as reviewed by Bonnell (2001).  
Separations of a few A.U. are possible however and further evolution of the binary can 
lead to short orbital period systems through the loss of angular momentum.
Therefore, it takes time for these systems to form.

It is widely known that young stars are rapid rotators and that main-sequence evolution 
causes a slowing of rotation.  The decrease of Ca II H\&K emission as a function of age 
in main sequence spectral types was clearly demonstrated by Wilson (1963).  
This leads to an activity versus age relationship amongst main-sequence stars as their 
stellar winds, which are coupled to the magnetic field and are forced to corotate with
stellar rotation out to large radii, extract angular momentum.
The result is that the stars rotate more  slowly over time.  
This in turn reduces the stellar activity level as speed of rotation is tied to magnetic dynamo strength.  
Since the result of extracting angular momentum from a binary system is to tighten the 
binary orbit, this angular momentum loss mechanism, when operating in a binary system, results 
in the evolution of the binary to shorter orbital periods. 
Stepien (1995) has formulated the angular momentum loss via a magnetized wind and the 
spin-up rate of binaries through their decreasing orbital periods calibrated from 
observations of the spin-down rates of single stars (Simon et al. 1985).  
These results show that the time-scales for angular momentum loss of rapidly rotating 
single stars increase substantially for cooler stars and later spectral types.  
Therefore, the time-scale for contact binary formation from initially more widely 
separated components will increase substantially with later spectral type components. 
Even the closest in ``situ'' formed binaries (2-5 A.U.), with low enough masses would not
have had enough time to reach the near-contact or contact stages within the age of our 
galaxy.  
Stepien's (1995) results show that for a binary with two 0.6 M$_\sun$ components, angular momentum loss
operating over 2.5 Gyr is needed to form a contact binary having an initial orbital period of 2 days.
For an initial orbital period of 3.4 days this increases to 6 Gyr.
The primary star of FS Aur-79 has a mass of approximately 0.6 M$_\sun$ based on its color and spectral type.
Our model indicates a mass ratio q of about 0.527, which results in a secondary mass of approximately 0.3 M$_\sun$, which
is consistent with a main-sequence star with the modeled temperature of 3425 K.
So the Stepien's (1995) results should also be roughly applicable to our system.
Given the main sequence life-times of our proposed components are $\approx$56 Gyr and the  age of the galactic disk is $\approx$12 Gyr (Kaluzny 1990), there has certainly been sufficient time for this binary to reach its current state from a wider initial separation.
Of the included binaries in the lists of near contact binaries (Shaw 1990, 1996), 
very few have spectral types later than F and none with M, making this system quite rare.

In fact, there are only a handful of stars thought to be of similar heritage.  
YY Gem and CM Dra (Strassmeier, et. al. 1988), CU Cnc (Ribas 2003), BW3 V38 (Maceroni \& Montalban 2004), 
and RE 0618+75 (Jeffries, et. al. 1993) are potential near-contact siblings along with 
their contact cousins, W Cor (Odell 1996), VZ Psc (Hrivnak et al. 1995),  for example, and those found listed by Hilditch et al. (1988), although none of those contact systems have a dMe component.

It is likely that the tidally enforced synchronous rotation of the stars in this system helps enhance their magnetic activity.  
Rotation, especially rapid rotation, is a prerequisite for the creation of strong magnetic 
phenomena seen in stars through an internal dynamo (for a recent review see Donati (2004)).  
FS  Aur-79 shows classic signs of chromospheric activity through the emission observed in 
H$\alpha$\ and H$\beta$\ from our spectroscopy (Figure 18) and star spots inferred from our multi-wavelength light 
curve modeling.  
It is therefore likely that this binary is evolving toward contact and has been for a long time.

The temperatures derived by light curve modeling (T$_1$=4100 K, T$_2$=3425 K) are supported by the 
observed spectra and are a good match (Figure 14) for a dK7e + dM3e near-contact system.  
The B-V color of the system also supports these spectral types.  
The U-B color would seem to be inconsistent with these spectral types indicating a 
hotter primary star.
However, because of the presumed star spots, the energy output in 
the UV is likely tainted by chromospheric activity, especially in the U-band (Spruit  \& Weiss, 1986; Houdebine, et. al. 1996).  
As an independent check of our temperature estimates, the depths of TiO absorption 
bands have been measured as described by O'Neal et al. (1998).  
In their paper they obtain empirical relationships between the depth of TiO absorption 
bands and temperature (their Figure 1).  
Measurements of the TiO bands at 7055\AA, 7088\AA, and 7127\AA\ in our high resolution spectra 
are consistent with our temperatures as shown in Figure 19 for TiO (7127\AA).

\section{Conclusions}

We find that FS Aur-79 is a near-contact binary system at a distance of $\approx$146 pc with dK7e and dM3e components separated by 1.62 R$_\sun$.
The system has a $\gamma$-velocity of +83$\pm$5.5 km sec$^{-1}$, with the primary having an orbital velocity around 115 km sec$^{-1}$, and
the secondary predicted to have an orbital velocity near 210 km sec$^{-1}$.
The primary mass, temperature, and radius are $\approx$0.6M$_\Sun$, 4100 K, and 0.67 R$_\Sun$. 
The secondary mass, temperature, radius are $\approx$ 0.3M$_\Sun$, 3425 K, and 0.48 R$_\Sun$.
This system is eclipsing with an inclination angle of 83 degrees, with unequal eclipse depths indicating 
the temperature difference, and unequal quadrature magnitudes indicating the presence of star spots.
A U-B excess, and H-Balmer lines in emission are further evidence of chromospheric activity.
The modeling indicates the presence of two major spot regions on the primary and one on the secondary.
It is likely that this system has evolved from larger separations through angular momentum loss via stellar
winds that are coupled to the magnetic fields from chromospheric activity and forced to corotate out to large radii
there by extracting angular momentum from the system and tightening the orbit.
This likely evolution means this system may become an W UMa system with unique late-type components, which is rare given that the evolution time is so long.

Further study of the high-resolution spectra are under way, including doppler-tomography, deconvolving
the primary and secondary spectra, and detailing of the components of emission features as a function of phase.
This should provide further information concerning the star spots, the chromospheric regions, and refinement of the stellar masses.

\acknowledgments

The authors acknowledge the observing time on the HET granted through NOAO.
S. J. Austin thanks the HET staff for their assistance.
J. W. Robertson would like to thank the members of the CBA, the everchanging vanguard at the IU RoboScope, and the time to take a spectrum at WIYN with the Hydra-MOS instrument. 
C.Tycner acknowledges support under a contract with the Jet Propulsion Laboratory (JPL) funded by NASA through the Michelson Fellowship Program. JPL is managed for NASA by the California Institute of Technology.  
C. T. would also like to thank Fred Vrba, Hugh Harris, Bob Zavala, and Trudy Tilleman for help
and assistance during his USNO 1-m telescope observing runs.

{\it Facilities:} \facility{USNO}, \facility{HET (HRS)}, \facility{WIYN}.

\clearpage

\begin{figure}
\plotone{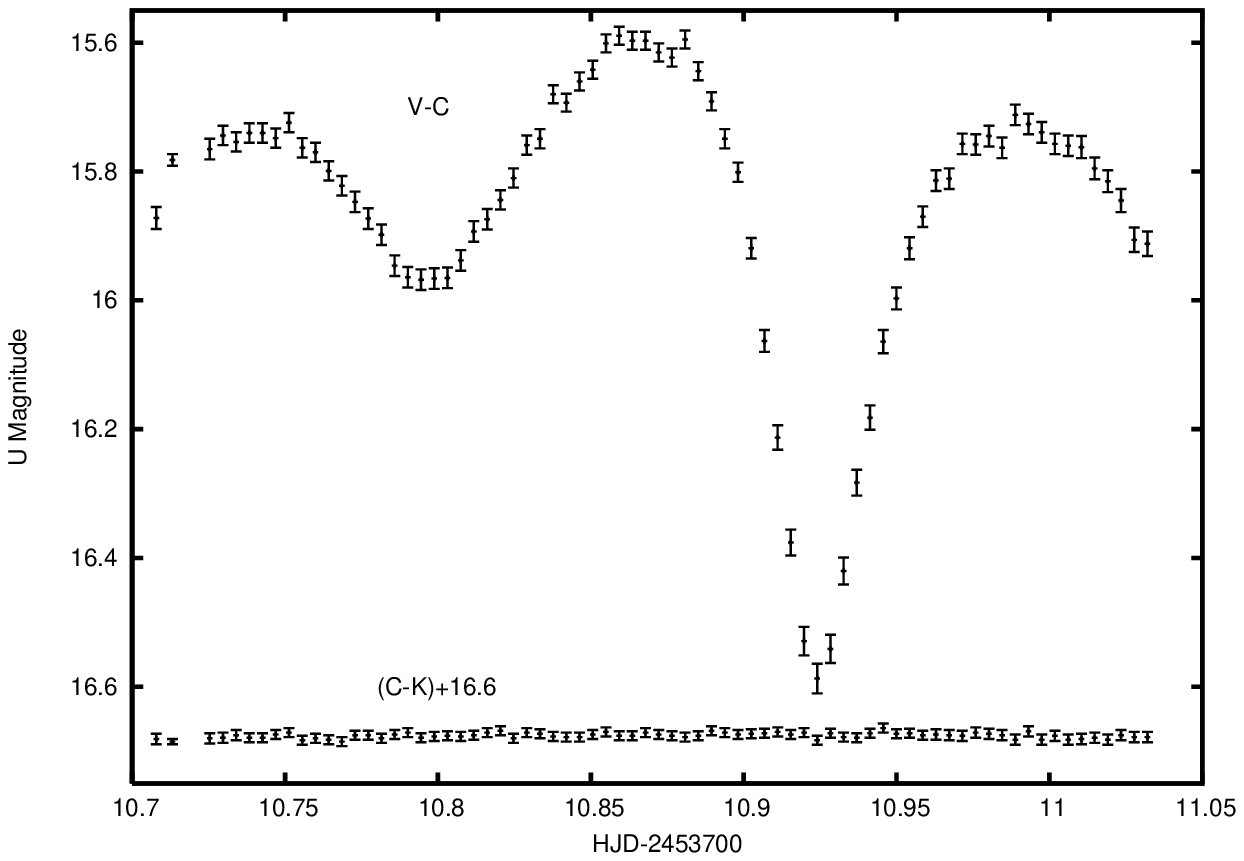}
\caption{U band photometry from 2005 December 6 obtained with the USNO 1-m telescope.  
The differential-U magnitudes have been converted to U magnitudes using the photometric 
standard stars for the FS Aur field.
The residual magnitudes of the C and K comparison stars are also shown.  \label{fig1}}
\end{figure}

\clearpage

\begin{figure}
\plotone{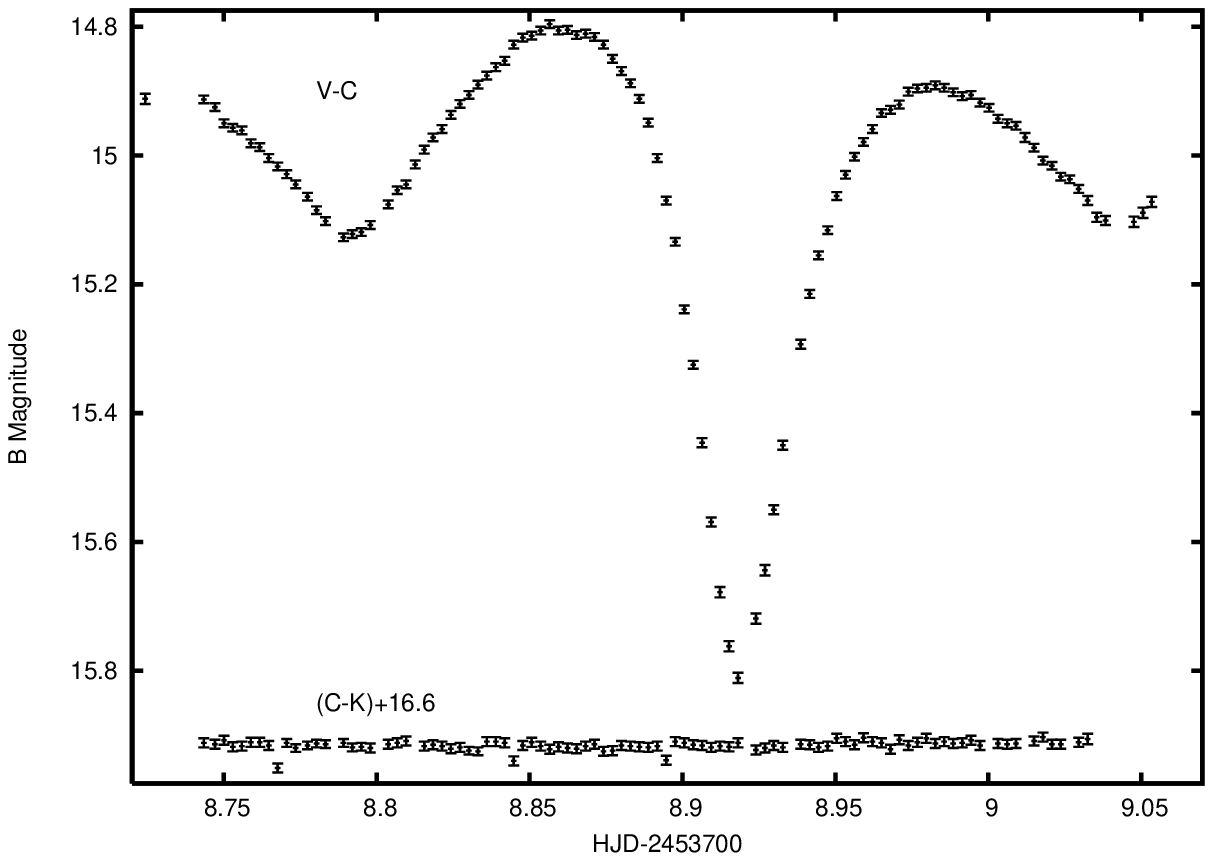}
\caption{B band photo photometry from 2005 December 4 obtained with the USNO 1-m telescope.
The differential-B magnitudes have been converted to B magnitudes using the photometric 
standard stars for the FS Aur field. 
The residual magnitudes of the C and K comparison stars are also shown.  \label{fig2}}
\end{figure}

\clearpage

\begin{figure}
\plotone{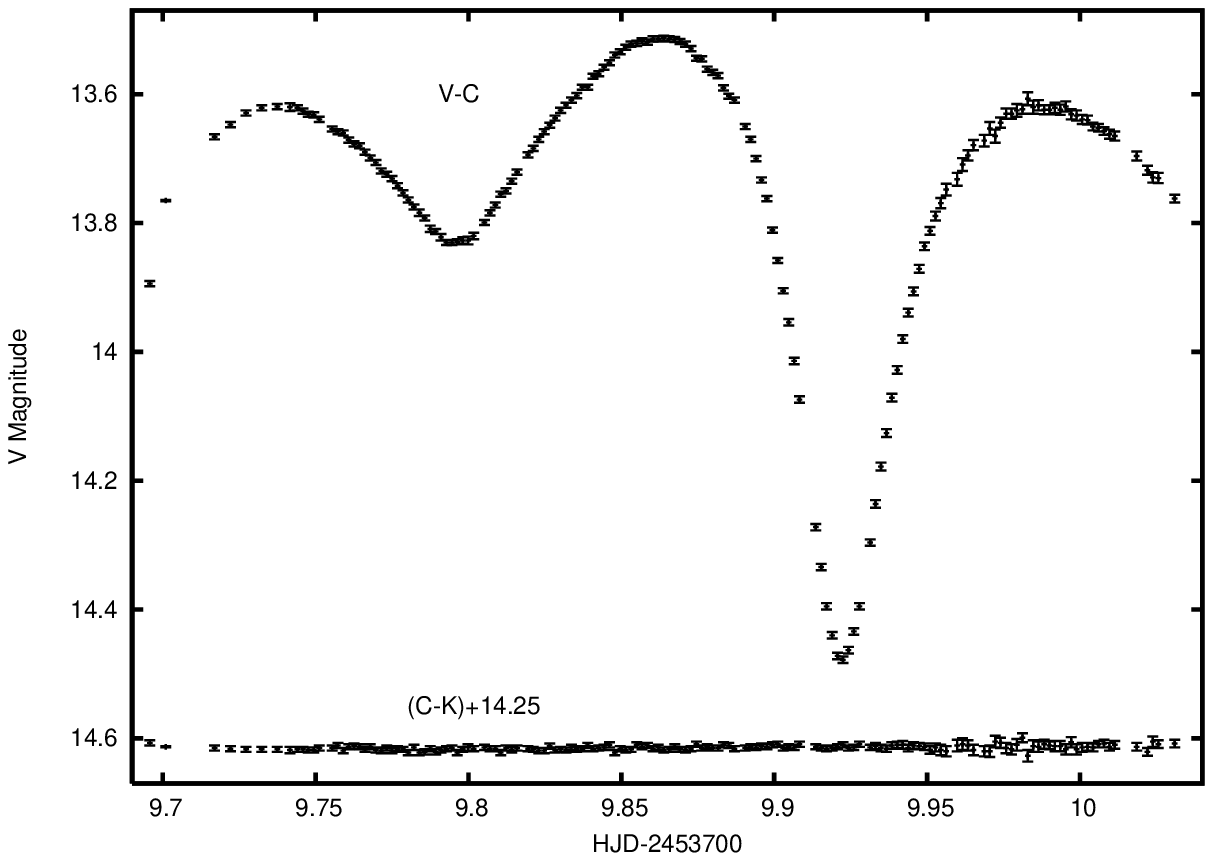}
\caption{V band photometry from 2005 December 5 obtained with the USNO 1-m telescope.
The differential-V magnitudes have been converted to V magnitudes using the photometric 
standard stars for the FS Aur field. 
The residual magnitudes of the C and K comparison stars are also shown.\label{fig3}}
\end{figure}

\clearpage

\begin{figure}
\plotone{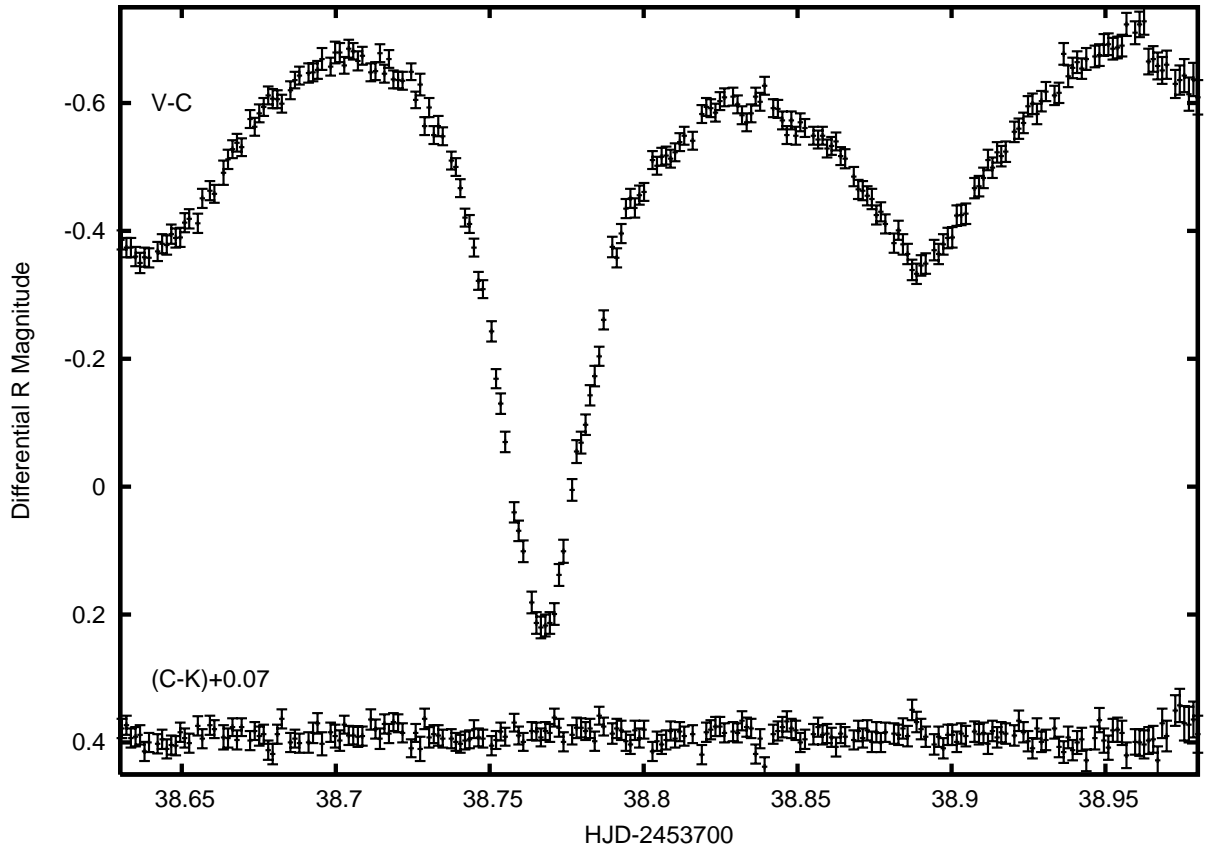}
\caption{R band photometry from 2006 January 2 obtained at Whispering Pines Observatory.
The differential-R magnitudes are shown, since comparison stars have not been calibrated for this
field.
The residual magnitudes of the C and K comparison stars is also shown.\label{fig4}}
\end{figure}

\clearpage

\begin{figure}
\plotone{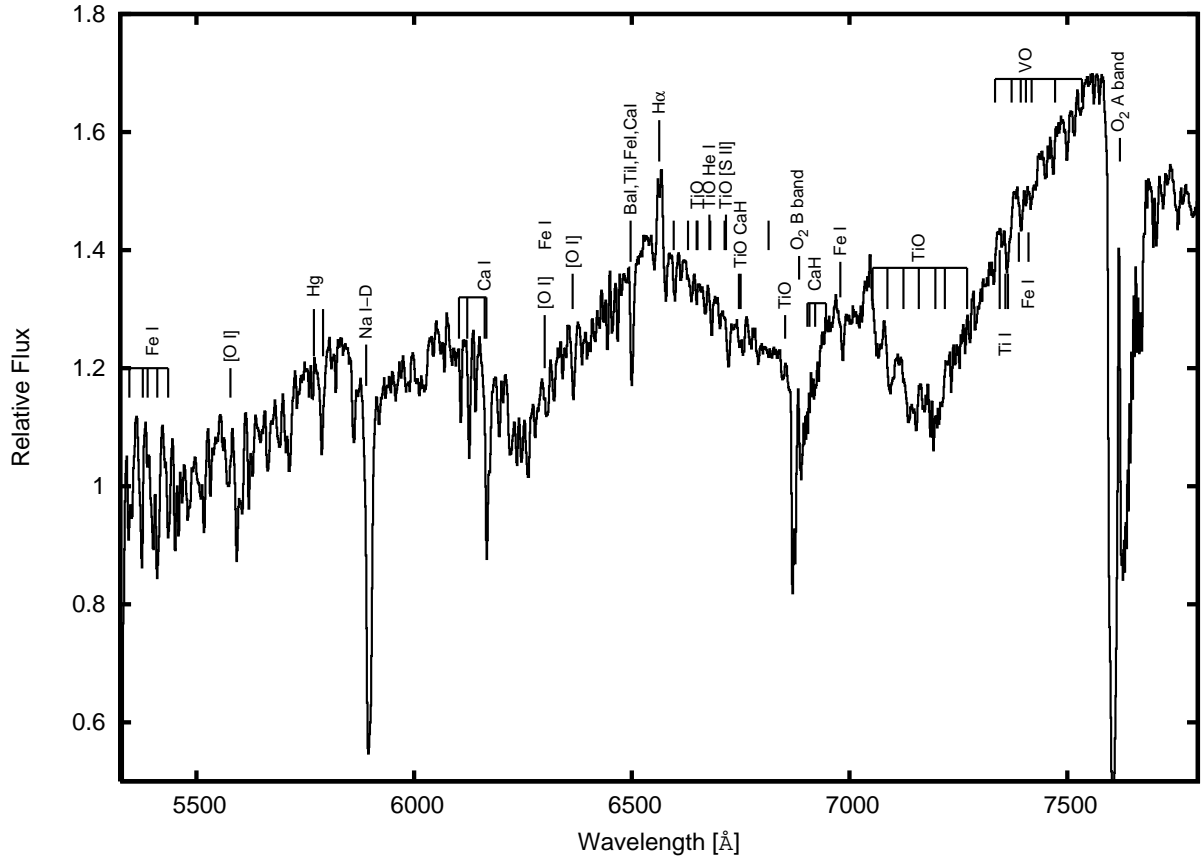}
\caption{The 1.4 \AA\ per pixel spectrum of FS Aur-79 from 2006 February 21 obtained with the WIYN telescope.
This spectrum was obtained during an orbital phase of about 0.76.
The major spectral features are labeled and are characteristic of K and M stars.\label{fig5}}
\end{figure}

\clearpage

\begin{figure}
\plotone{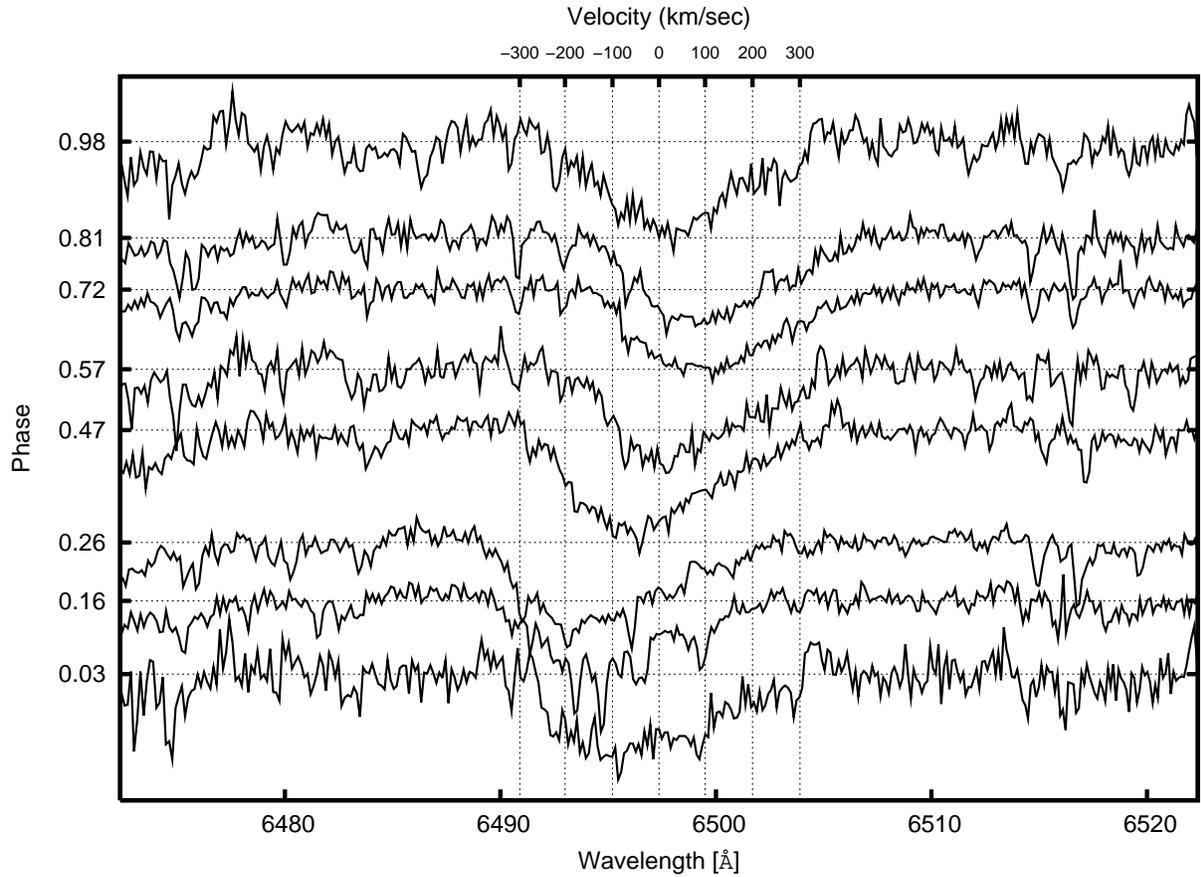}
\caption{Spectra of the BaI-TiI-FeI-CaI blend at representative orbital phases obtained with the HET HRS 
(Order 94) normalized to the continuum (horizontal dotted lines).
The velocity scale is for the center-of-mass frame of the binary.\label{fig6}}
\end{figure}

\clearpage

\begin{figure}
\plotone{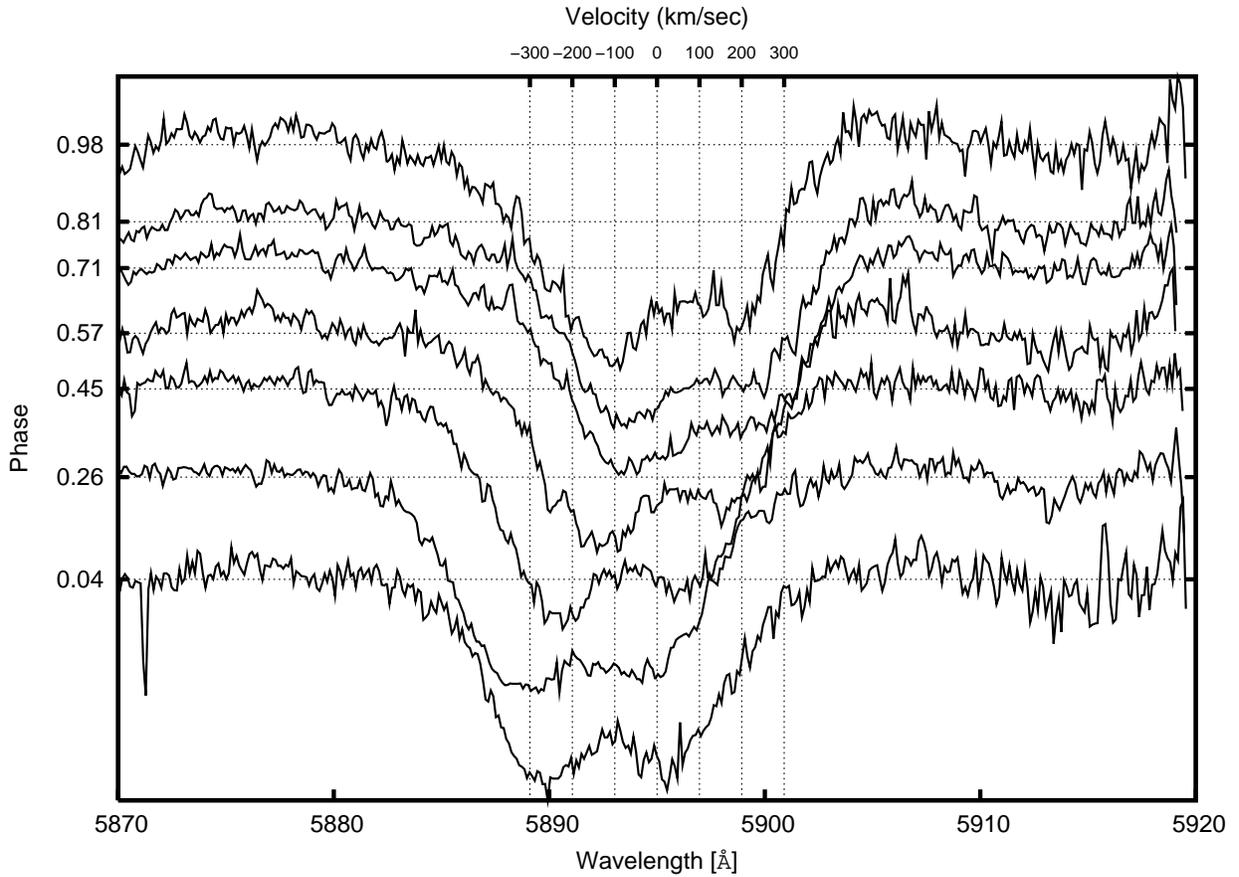}
\caption{Spectra of the NaI-D doublet at representative orbital phases obtained with the HET HRS 
(Order 104) normalized to the continuum (horizontal dotted lines).
The velocity scale is for the center-of-mass frame of the binary.\label{fig7}}
\end{figure}

\clearpage

\begin{figure}
\plotone{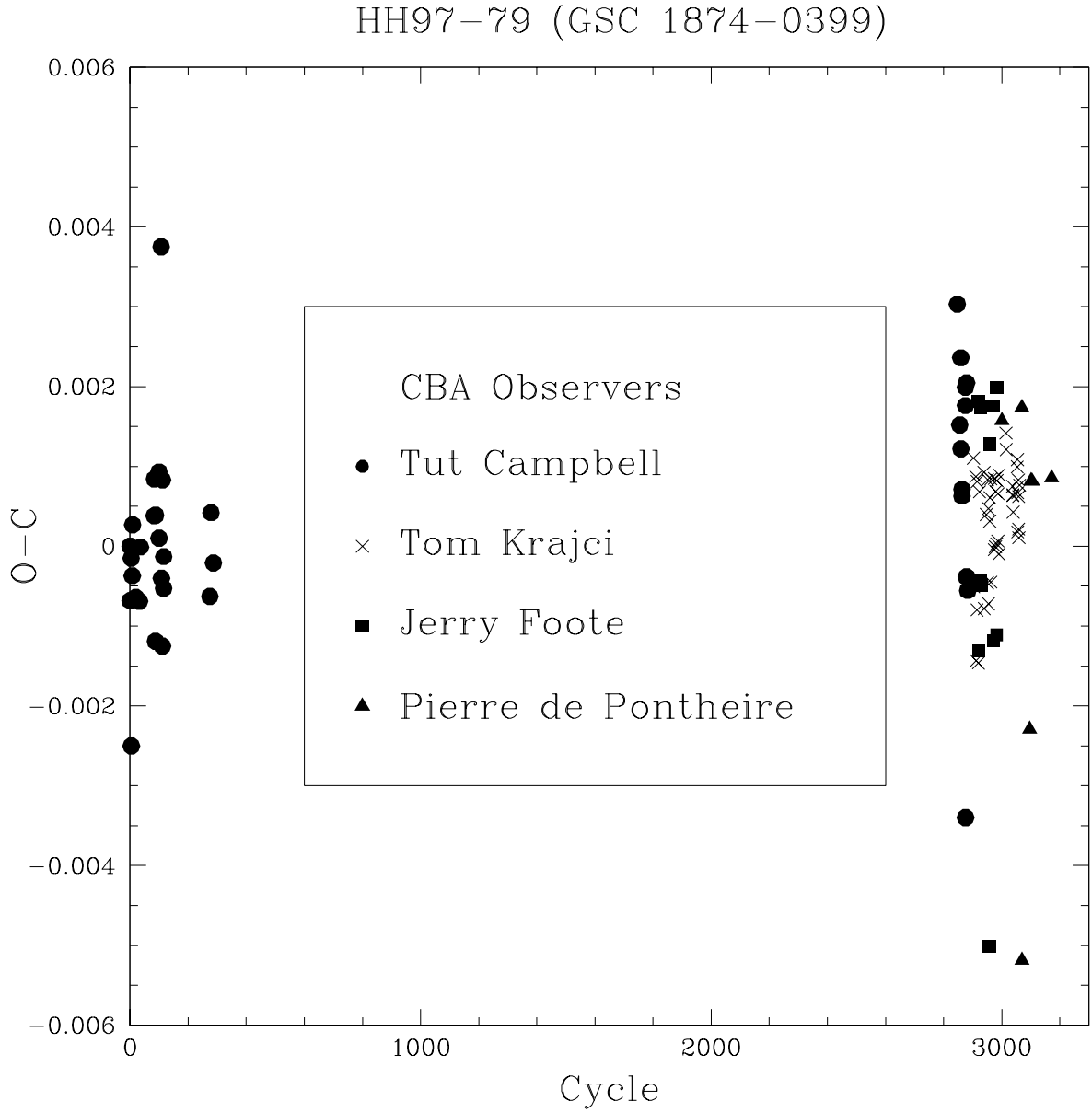}
\caption{Difference between observed and calculated (O-C) eclipse timings from CCD photometry
obtained by members of the CBA.\label{fig8}}
\end{figure}

\clearpage

\begin{figure}
\plotone{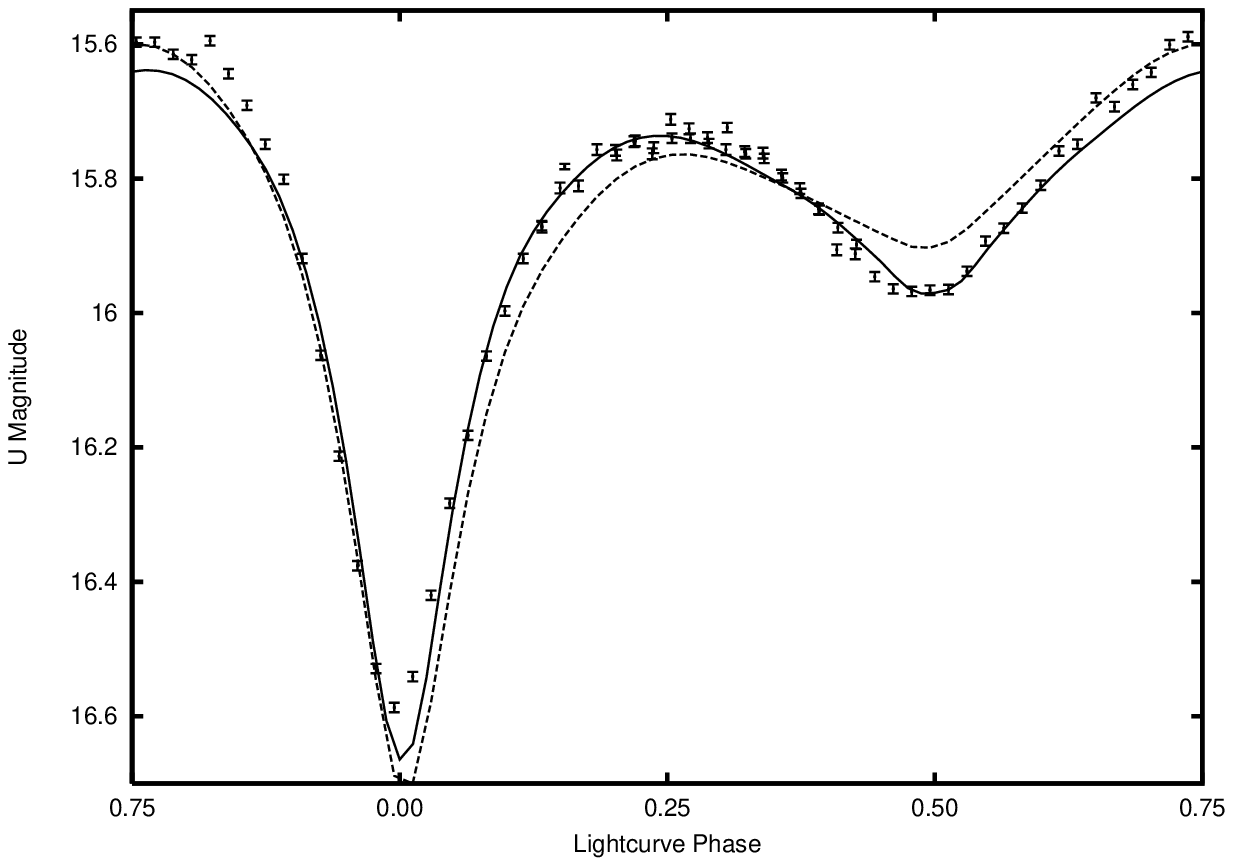}
\caption{Phased U band photometry from 2005 December 6 obtained with the USNO 1-m telescope.
Solid curve is the best-fit {\tt Nightfall} model.
Dashed curve is the {\tt BM3} model. \label{fig9}}
\end{figure}

\clearpage

\begin{figure}
\plotone{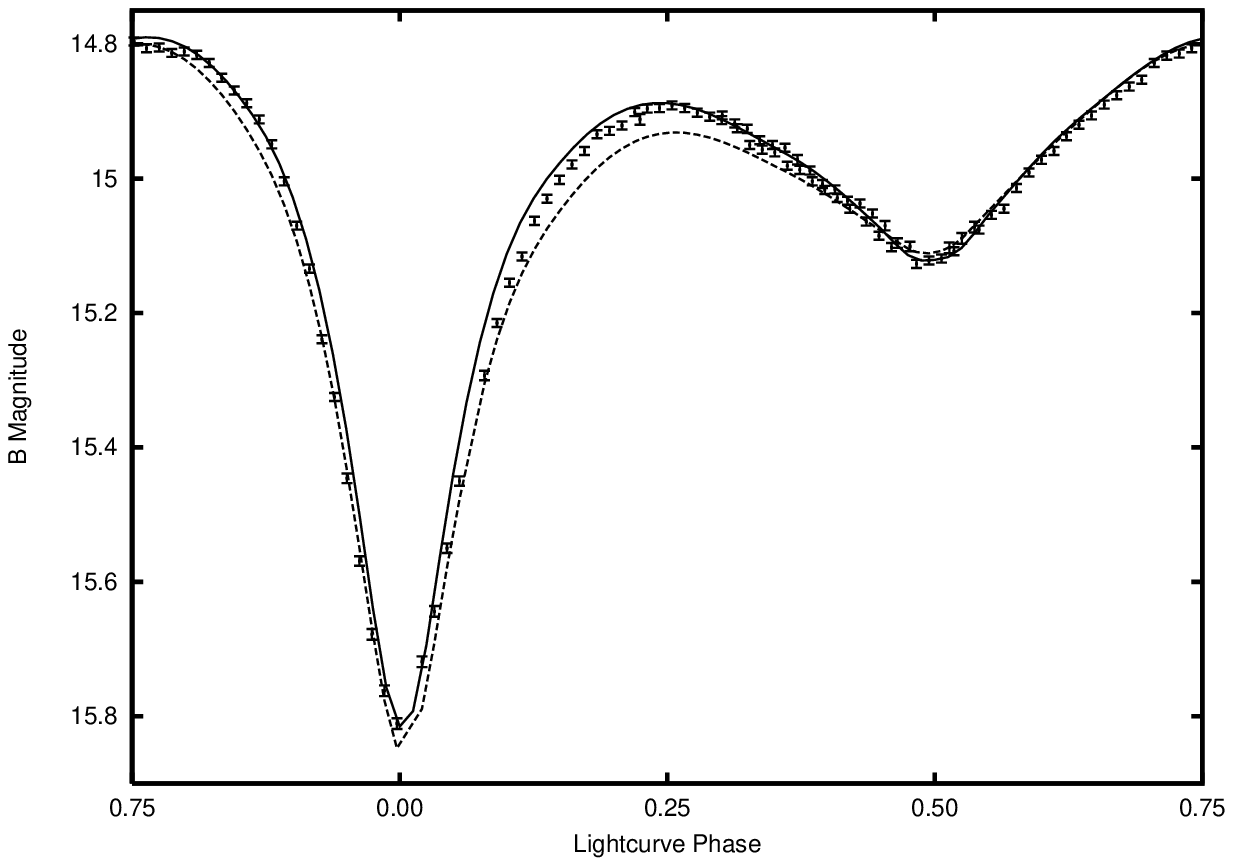}
\caption{Phased B band photometry from 2005 December 4 obtained with the USNO 1-m telescope.
Solid curve is the best-fit {\tt Nightfall} model.
Dashed curve is the {\tt BM3} model. \label{fig10}}
\end{figure}

\clearpage

\begin{figure}
\plotone{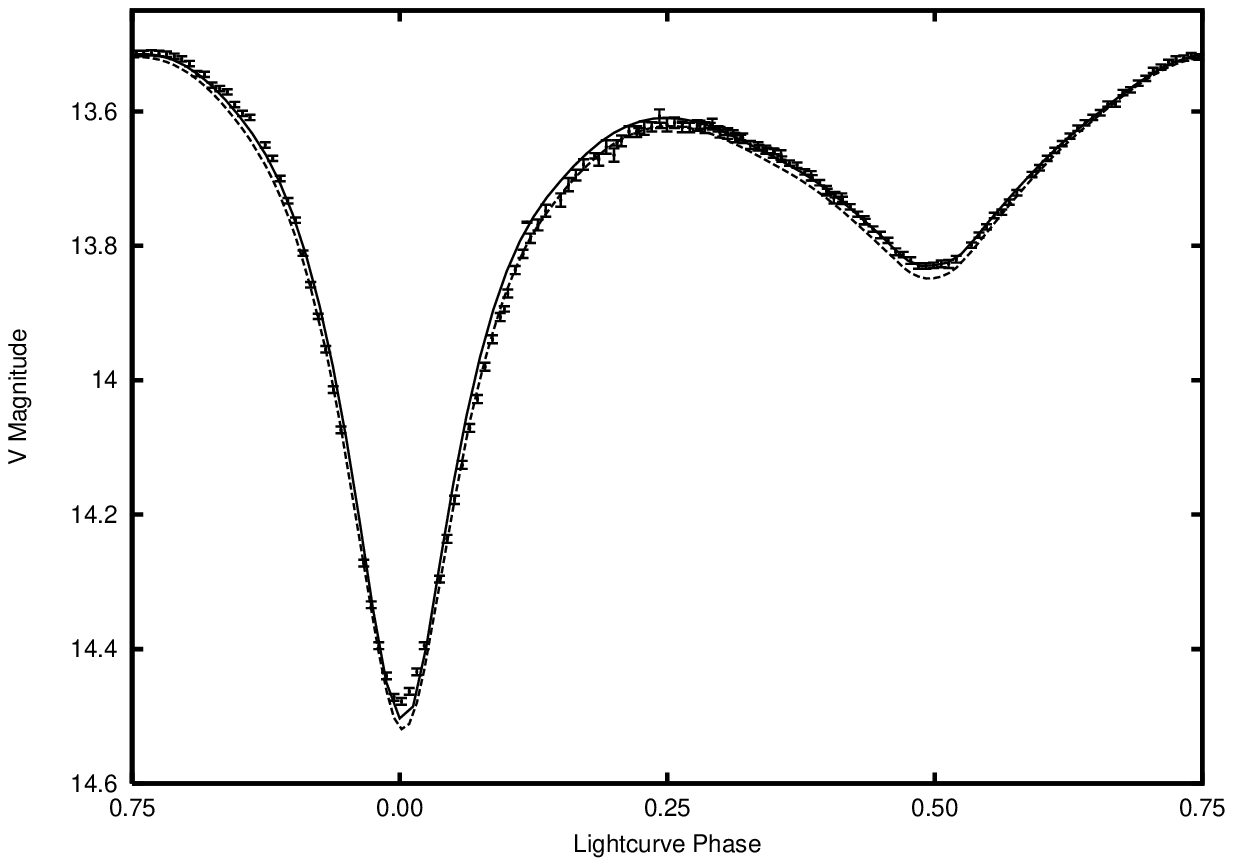}
\caption{Phased V band photometry from 2005 December 5 obtained with the USNO 1-m telescope.
Solid curve is the best-fit {\tt Nightfall} model.
Dashed curve is the {\tt BM3} model.  \label{fig11}}
\end{figure}

\clearpage

\begin{figure}
\plotone{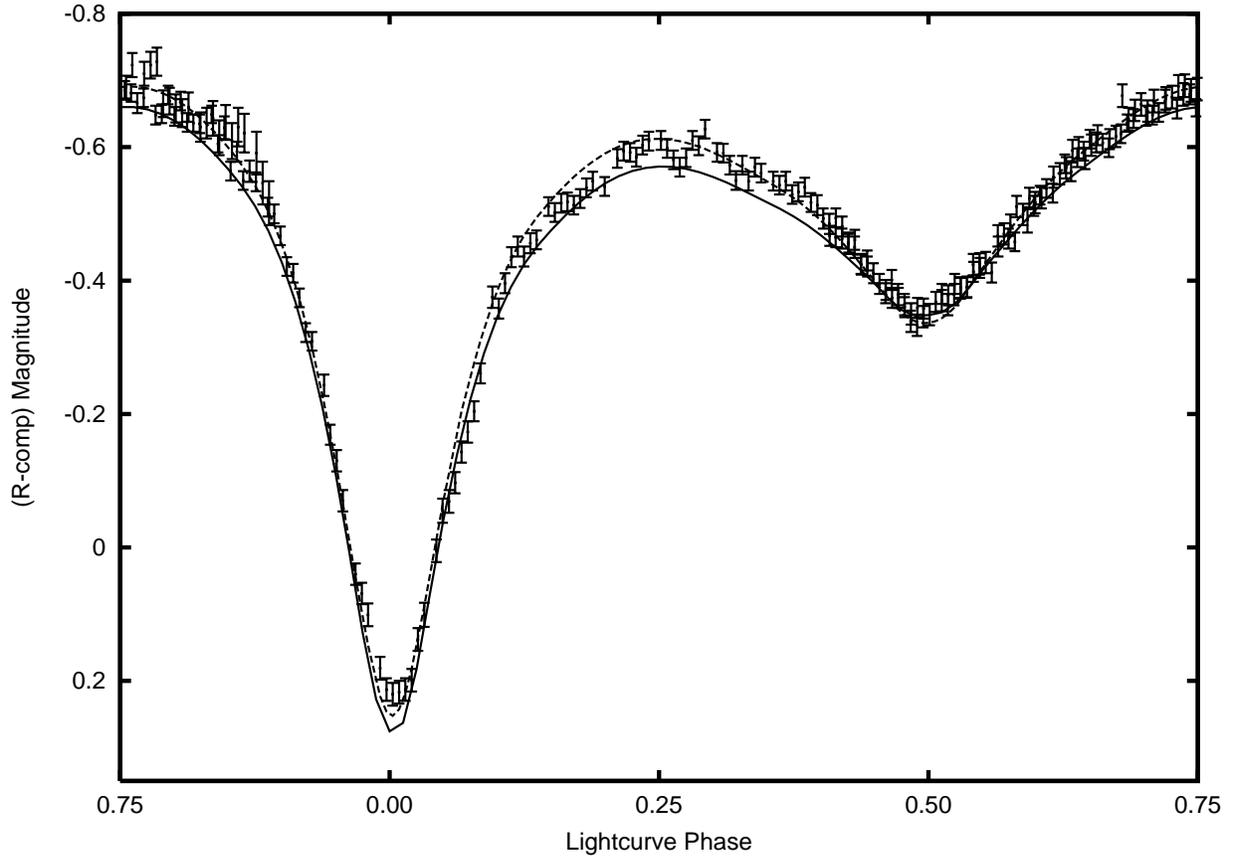}
\caption{Phased R band photometry from 2006 January 2 obtained at Whispering Pines Observatory.
Solid curve is the best-fit {\tt Nightfall} model.
Dashed curve is the {\tt BM3} model.  \label{fig12}}
\end{figure}

\clearpage

\begin{figure}
\plotone{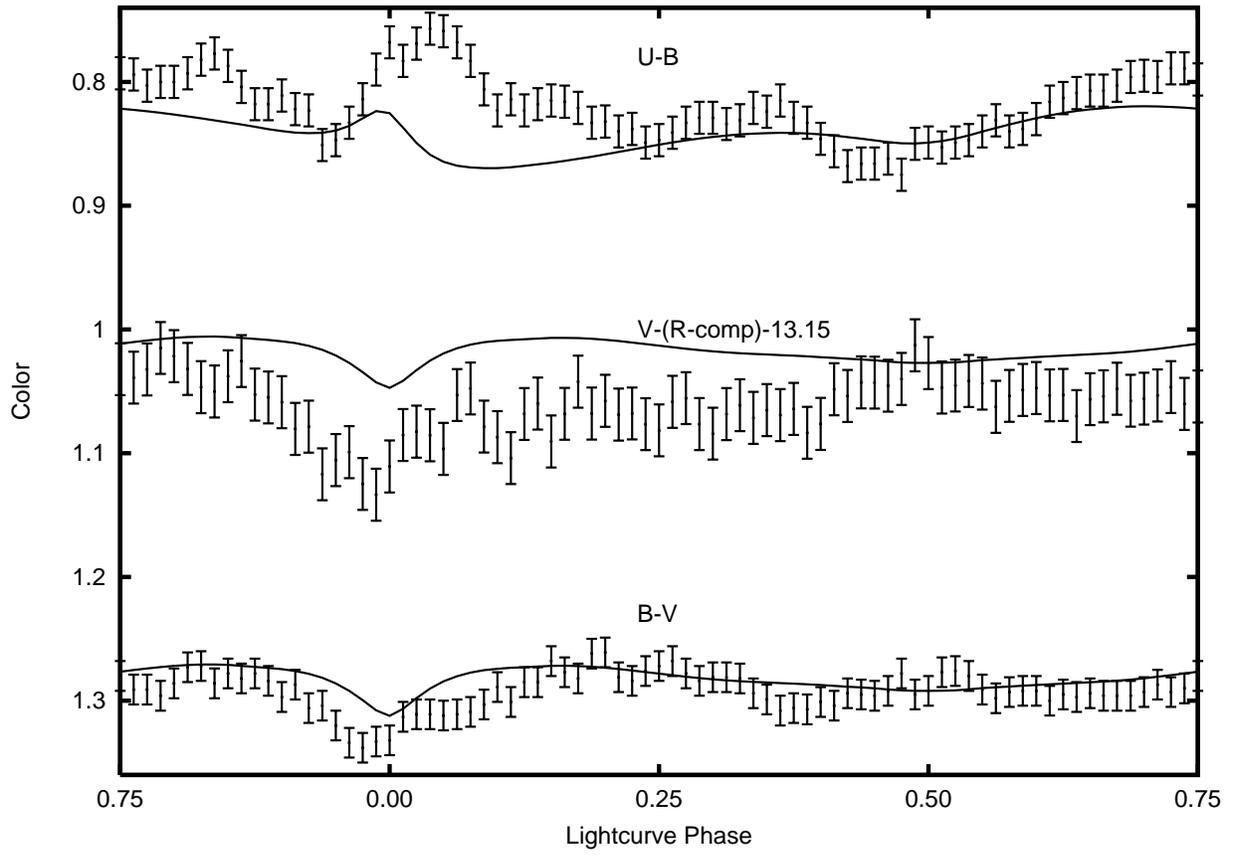}
\caption{Phased U-B, B-V, and V-(R-comp) light curve.
Solid lines are from the best-fit {\tt Nightfall} model.\label{fig13}}
\end{figure}

\clearpage

\begin{figure}
\plotone{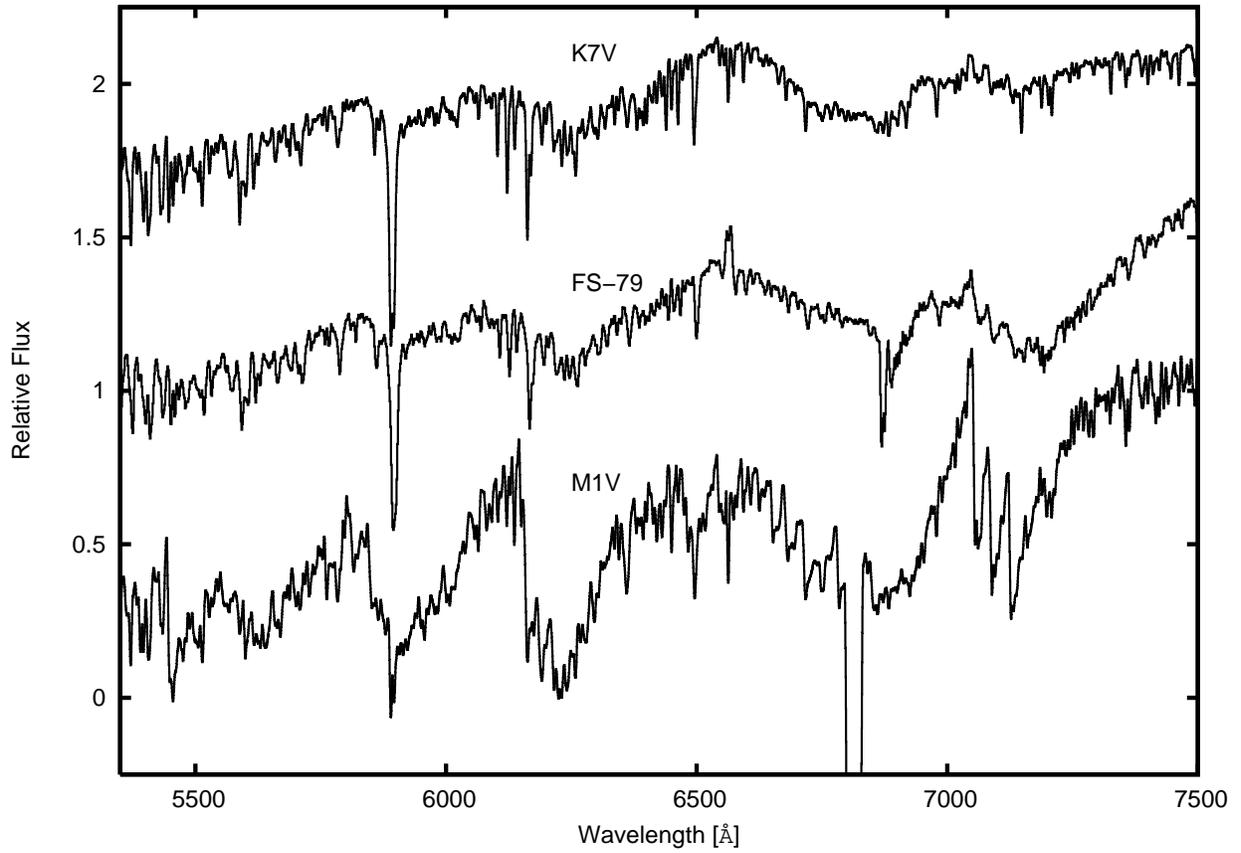}
\caption{The 1.4 \AA\ per pixel spectrum of FS Aur-79 from 2006 February 21 obtained with the WIYN telescope compared to rotationally broadened (vsin(i)=135 km sec$^{-1}$) spectra of K7V and M1V stars (Indo-US Library stars HD237903 and HD204445 (Valdes et al. 2004))  \label{fig14}}
\end{figure}

\clearpage

\begin{figure}
\plotone{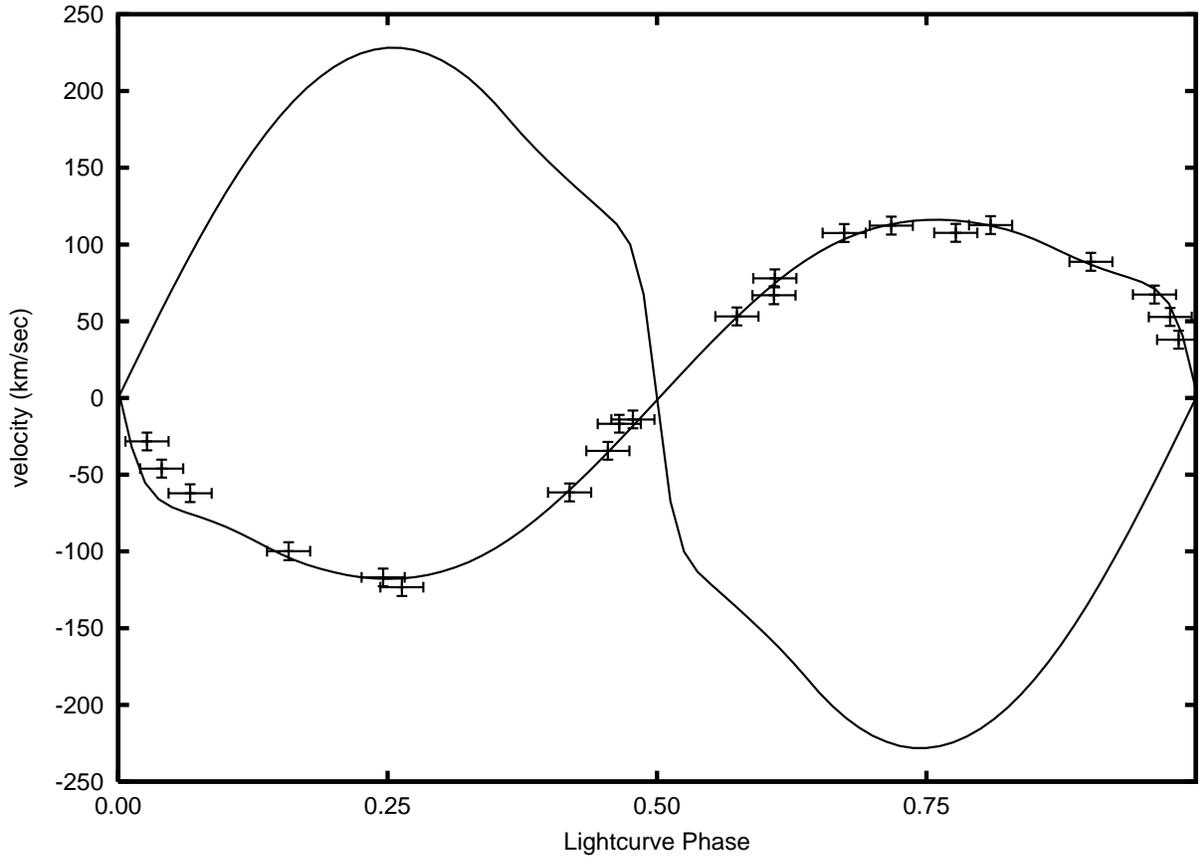}
\caption{Radial velocity curve of the primary star.
Solid curves are the model radial velocity curves for the primary and secondary.\label{fig15}}
\end{figure}

\clearpage

\begin{figure}
\plotone{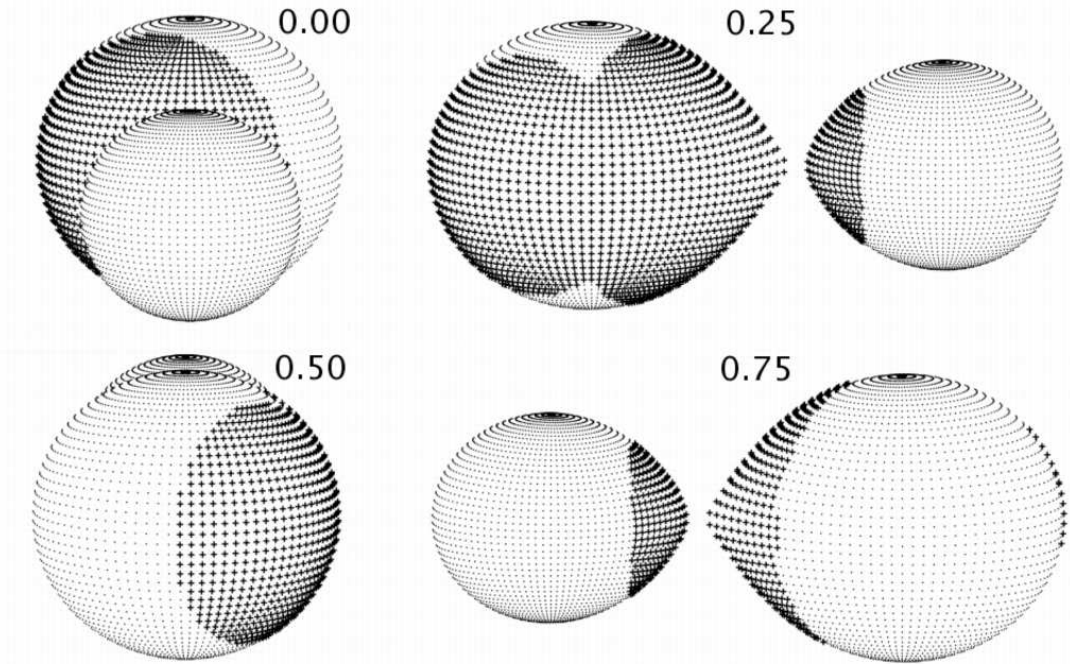}
\caption{Configuration of spots for the best-fit three-spot binary model shown at orbital phases 0.00, 0.25, 0.50, and 0.75.. \label{fig16}}
\end{figure}

\clearpage

\begin{figure}
\plotone{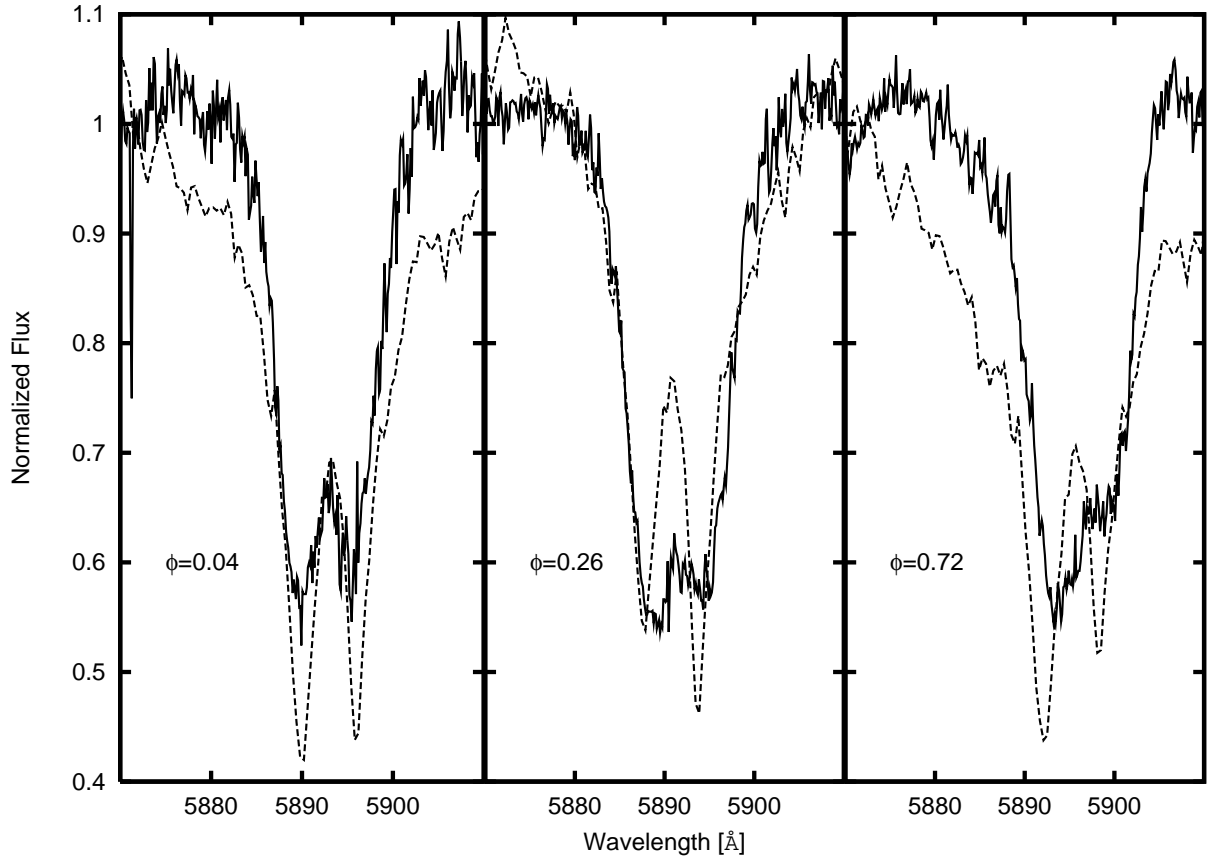}
\caption{Sodium doublet and synthesized profiles for phases 0.0, 0.25, 0.5, 0.75.\label{fig17}}
\end{figure}

\clearpage

\begin{figure}
\plotone{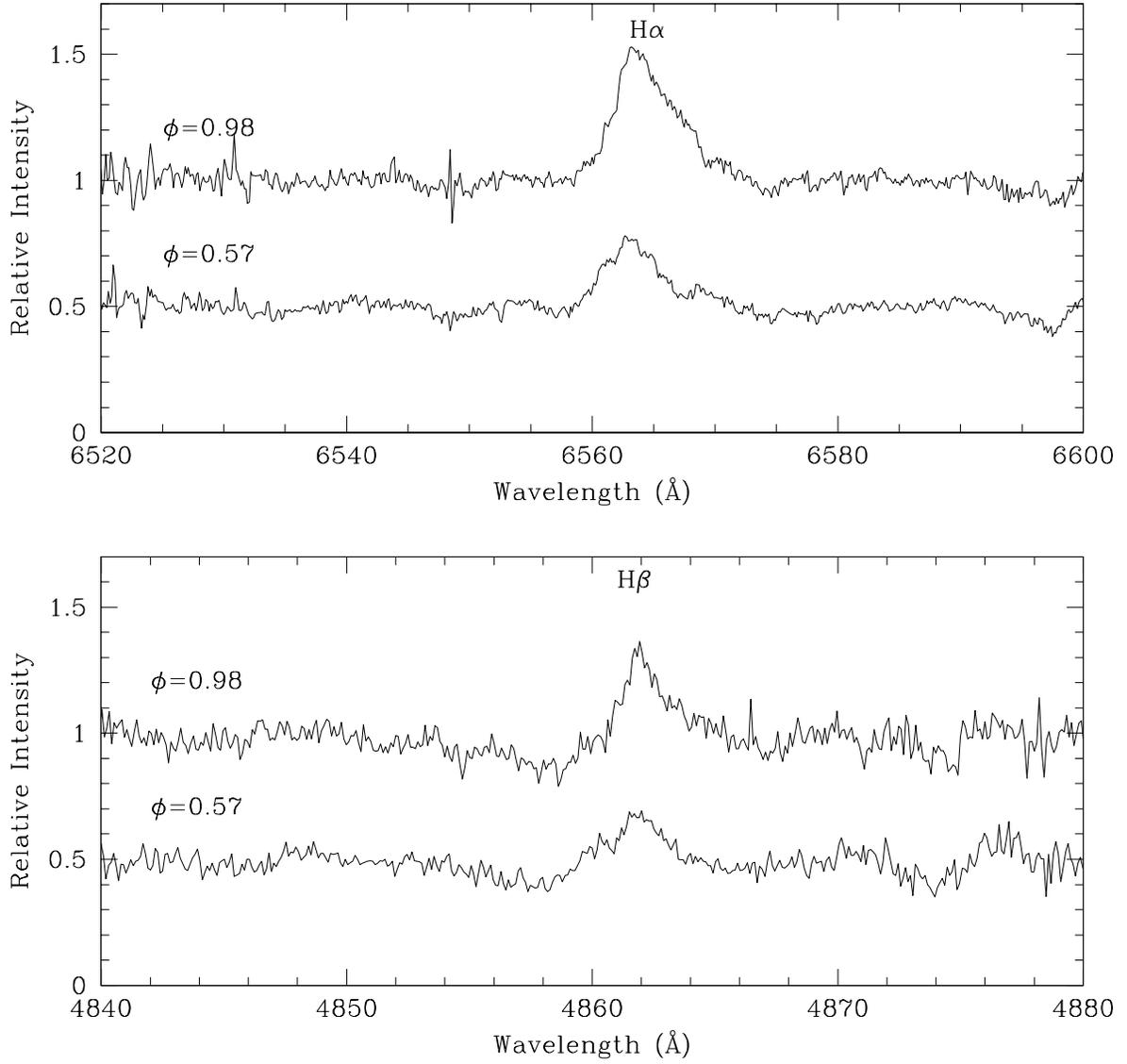}
\caption{Examples of H$\alpha$ and H$\beta$ in emission lines at phases 0.0 and 0.5.\label{fig18}}
\end{figure}

\begin{figure}
\plotone{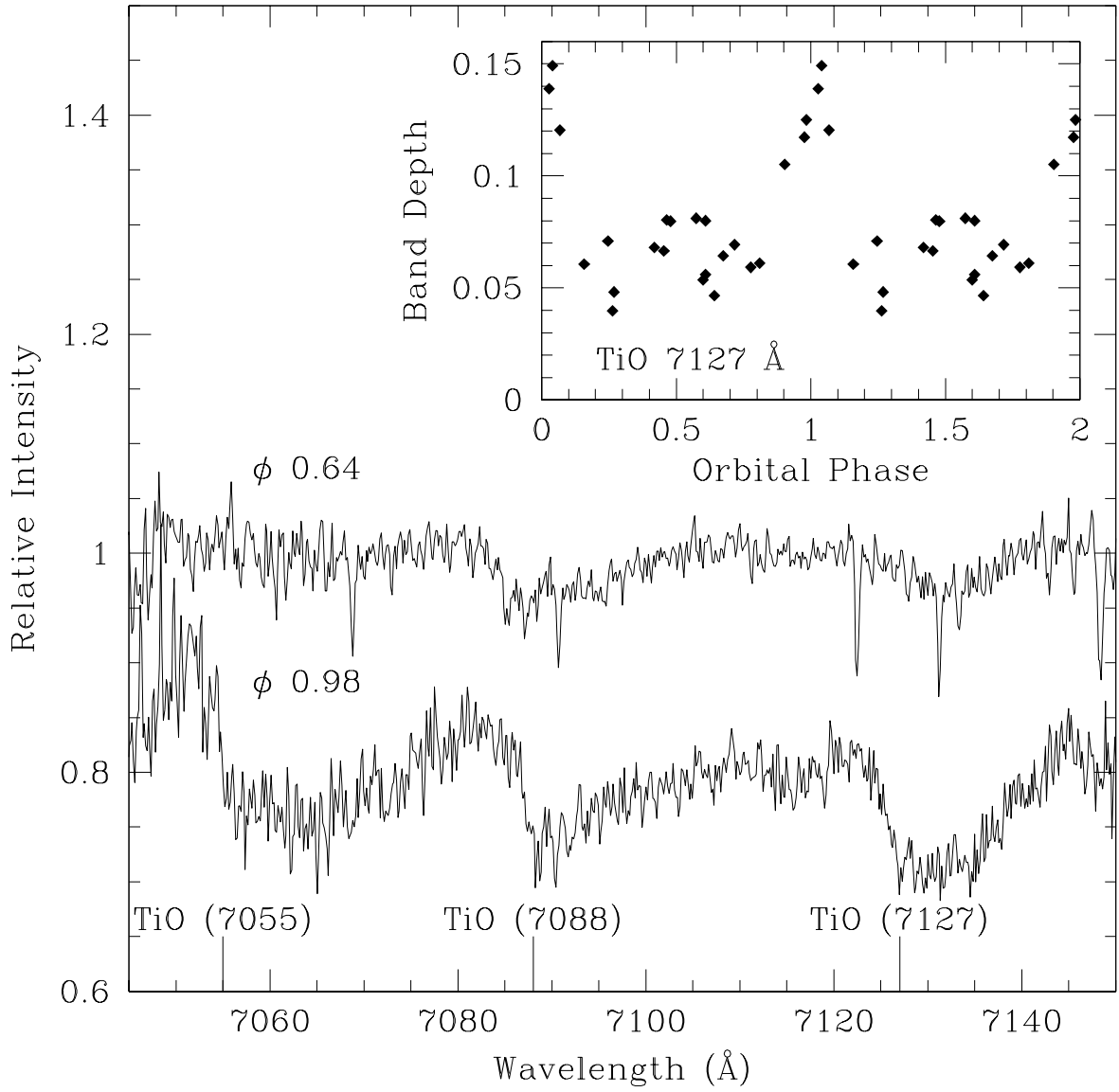}
\caption{TiO bands and band depths.\label{fig19}}
\end{figure}

\clearpage

\begin{deluxetable}{cccc}
\tablewidth{0pt}
\tablecaption{CBA Times of Minima}
\tablehead{
\colhead{HJD} & \colhead{Phase Cycle}&  \colhead{Filter} & \colhead{Observatory}
}
\startdata
2452963.74445 & 0.0 & R & CBA-Arkansas \\
2452963.86918 & 0.5 & R &CBA-Arkansas \\
2452964.74756 & 4.0 & R & CBA-Arkansas \\
2452964.87062 & 4.5 & R & CBA-Arkansas \\
2452965.75061 & 8.0 & R & CBA-Arkansas \\
2452966.00206 & 9.0 & R & CBA-Arkansas \\
2452968.76013 & 20.0 & R & CBA-Arkansas \\
2452971.89528 & 32.5 & R & CBA-Arkansas \\
2452972.77382 & 36.0 & R & CBA-Arkansas \\
2452984.68797 & 83.5 & V & CBA-Arkansas \\
2452984.81383 & 84.0 & V & CBA-Arkansas \\
2452985.68966 & 87.5 & V & CBA-Arkansas \\
2452985.81665 & 88.0 & V & CBA-Arkansas \\
2452988.70157 & 99.5 & V & CBA-Arkansas \\
2452988.82615 & 100.0 & V & CBA-Arkansas \\
2452990.71092 & 107.5 & V & CBA-Arkansas \\
2452990.83218 & 108.0 & V & CBA-Arkansas \\
2452991.70918 & 111.5 & V & CBA-Arkansas \\
2452991.83667 & 112.0 & V & CBA-Arkansas \\
2452992.71317 & 115.5 & V & CBA-Arkansas \\
2452992.83898 & 116.0 & V & CBA-Arkansas \\
2453032.71822 & 275.0 & B & CBA-Arkansas \\
2453033.72253 & 279.0 & B & CBA-Arkansas \\
2453035.72843 & 287.0 & B & CBA-Arkansas \\
2453691.86440 & 2903.0 & none & CBA-N.M.\tablenotemark{a}\\
2453691.98820 & 2903.5 & none & CBA-N.M.\tablenotemark{a}\\
2453693.87070 & 2911.0 & none & CBA-N.M.\tablenotemark{a}\\
2453693.99380 & 2911.5 & none & CBA-N.M.\tablenotemark{a}\\
2453694.87390 & 2915.0 & none & CBA-N.M.\tablenotemark{a}\\
2453694.99770 & 2915.5 & none & CBA-N.M.\tablenotemark{a}\\
2453695.75052 & 2918.5 & none & CBA-Utah\\
2453695.87817 & 2919.0 & none & CBA-Utah\\
2453696.00030 & 2919.5 & none & CBA-N.M.\tablenotemark{a}\\
2453696.00045 & 2919.5 & none & CBA-Utah\\
2453696.88030 & 2923.0 & none & CBA-N.M.\tablenotemark{a}\\
2453697.75706 & 2926.5 & none & CBA-Utah\\
2453697.88463 & 2927.0 & none & CBA-Utah\\
2453698.00781 & 2927.5 & none & CBA-Utah\\
2453700.89360 & 2939.0 & none & CBA-N.M.\tablenotemark{a}\\
2453701.01730 & 2939.5 & none & CBA-N.M.\tablenotemark{a}\\
2453702.89960 & 2947.0 & none & CBA-N.M.\tablenotemark{a}\\
2453703.77660 & 2950.5 & none & CBA-N.M.\tablenotemark{a}\\
2453703.90330 & 2951.0 & none & CBA-N.M.\tablenotemark{a}\\
2453704.77960 & 2954.5 & none & CBA-N.M.\tablenotemark{a}\\
2453704.90620 & 2955.0 & none & CBA-N.M.\tablenotemark{a}\\
2453705.77857 & 2958.5 & none & CBA-Utah\\
2453705.78390 & 2958.5 & none & CBA-N.M.\tablenotemark{a}\\
2453705.90960 & 2959.0 & none & CBA-N.M.\tablenotemark{a}\\
2453705.91027 & 2959.0 & none & CBA-Utah\\
2453706.78640 & 2962.5 & none & CBA-N.M.\tablenotemark{a}\\
2453706.91310 & 2963.0 & none & CBA-N.M.\tablenotemark{a}\\
2453708.79220 & 2970.5 & none & CBA-Utah\\
2453708.92054 & 2971.0 & none & CBA-Utah\\
2453709.79660 & 2974.5 & none & CBA-N.M.\tablenotemark{a}\\
2453709.92290 & 2975.0 & none & CBA-N.M.\tablenotemark{a}\\
2453710.79990 & 2978.5 & none & CBA-N.M.\tablenotemark{a}\\
2453710.92600 & 2979.0 & none & CBA-N.M.\tablenotemark{a}\\
2453711.80206 & 2982.5 & none & CBA-Utah\\
2453711.80320 & 2982.5 & none & CBA-N.M.\tablenotemark{a}\\
2453711.92940 & 2983.0 & none & CBA-N.M.\tablenotemark{a}\\
2453711.93056 & 2983.0 & none & CBA-Utah\\
2453712.80650 & 2986.5 & none & CBA-N.M.\tablenotemark{a}\\
2453712.93250 & 2987.0 & none & CBA-N.M.\tablenotemark{a}\\
2453713.80960 & 2990.5 & none & CBA-N.M.\tablenotemark{a}\\
2453713.93600 & 2991.0 & none & CBA-N.M.\tablenotemark{a}\\
2453716.44484 & 3001.0 & none & CBA-Lesve\\
2453719.83070 & 3014.5 & none & CBA-N.M.\tablenotemark{a}\\
2453719.95590 & 3015.0 & none & CBA-N.M.\tablenotemark{a}\\
2453725.59880 & 3037.5 & none & CBA-N.M.\tablenotemark{a}\\
2453725.72410 & 3038.0 & none & CBA-N.M.\tablenotemark{a}\\
2453725.84950 & 3038.5 & none & CBA-N.M.\tablenotemark{a}\\
2453725.97470 & 3039.0 & none & CBA-N.M.\tablenotemark{a}\\
2453729.61210 & 3053.5 & none & CBA-N.M.\tablenotemark{a}\\
2453729.73760 & 3054.0 & none & CBA-N.M.\tablenotemark{a}\\
2453729.86210 & 3054.5 & none & CBA-N.M.\tablenotemark{a}\\
2453729.98800 & 3055.0 & none & CBA-N.M.\tablenotemark{a}\\
2453730.61520 & 3057.5 & none & CBA-N.M.\tablenotemark{a}\\
2453730.74040 & 3058.0 & none & CBA-N.M.\tablenotemark{a}\\
2453730.86540 & 3058.5 & none & CBA-N.M.\tablenotemark{a}\\
2453730.99070 & 3059.0 & none & CBA-N.M.\tablenotemark{a}\\
2453731.61840 & 3061.5 & none & CBA-N.M.\tablenotemark{a}\\
2453733.50049 & 3069.0 & none & CBA-Lesve\\
2453733.61898 & 3069.5 & none & CBA-Lesve\\
2453740.39391 & 3096.5 & none & CBA-Lesve\\
2453741.52569 & 3101.0 & none & CBA-Lesve\\
2453742.52894 & 3105.0 & none & CBA-Lesve\\
2453759.33366 & 3172.0 & none & CBA-Lesve\\
2453677.82063 & 2847.0 & V &  CBA-Arkansas \\
2453677.95151 & 2847.5 & V &  CBA-Arkansas \\
2453679.951055 & 2855.5 & V &  CBA-Arkansas \\
2453680.82975 & 2859.0 & B & CBA-Arkansas \\
2453680.95402 & 2859.5 & V & CBA-Arkansas \\
2453681.83129 & 2863.0 & B & CBA-Arkansas \\
2453681.83137 & 2863.0 & V &  CBA-Arkansas \\
2453684.84244 & 2875.0 & B & CBA-Arkansas \\
2453684.84221 & 2875.0 & V & CBA-Arkansas \\
2453684.96246 & 2875.5 & V &  CBA-Arkansas \\
2453685.84576 & 2879.0 & B & CBA-Arkansas \\
2453685.96874 & 2879.5 & V & CBA-Arkansas \\
2453685.84333 & 2879.0 & V & CBA-Arkansas \\
2453686.84642 & 2883.0 & V & CBA-Arkansas \\
\enddata
\tablenotetext{a}{Previously published in IBVS 5690}
\end{deluxetable}

\clearpage

\begin{deluxetable}{cc}
\tablecaption{WIYN Observations}
\tablewidth{0pt}
\tablehead{
\colhead{HJD} & \colhead{Orbital Phase}
}
\startdata
2453788.6109458 & 0.732\\
2453788.6194817 & 0.766\\
2453788.6283672 & 0.801\\
\enddata
\end{deluxetable}

\clearpage

\begin{deluxetable}{cccccc}
\tablecaption{HET Observations}
\tablewidth{0pt}
\tablehead{
\colhead{UT Date} & \colhead{UT Time} & \colhead{Orbital Phase} & \colhead{Exp (sec)} & \colhead{Sky\tablenotemark{a}} & \colhead{S/N (per res)}
}
\startdata
2005-10-09   & 09:09 & 0.609 & 980 & S & 75\\
2005-10-12   & 08:31 & 0.465& 900 & P & 124\\
2005-10-20   & 07:48 & 0.246& 900 & P & 40\\
2005-11-01   & 07:29 & 0.040& 900 & P & 78\\
2005-11-03   & 07:15 & 0.976& 900 & P & 82\\
2005-11-11   & 06:40 & 0.777& 900 & P & 150\\
2005-11-11   & 11:40 & 0.609& 900 & P & 135\\
2005-11-19   & 06:13 & 0.599& 900 & P & 140\\
2005-11-19   & 11:22 & 0.454& 900 & P & 137\\
2005-11-26   & 05:40 & 0.419& 900 & S & 131\\
2005-11-26   & 10:45 & 0.263& 900 & P & 153\\
2005-12-10   & 10:01 & 0.962& 900 & P & 120\\
2005-12-19   & 03:47 & 0.810& 1000 &P  & 135\\
2005-12-19   & 09:15 & 0.717& 900 & P & 161\\
2005-12-27   & 03:36 & 0.674& 900 & S & 114\\
2005-12-28   & 09:06 & 0.574& 900 & S & 93\\
2006-01-03   & 02:58 & 0.478& 900 & P & 114\\
2006-01-11   & 02:21 & 0.269& 900 & S & 110\\
2006-01-11   & 07:42 & 0.158& 900 & S & 130\\
2006-01-19   & 07:34 & 0.030& 900 & S & \nodata \\
2006-01-20   & 01:51 & 0.067& 900 & P & 90\\
2006-01-21   & 07:09 & 0.935& 900 & P & \nodata \\
2006-01-21   & 07:27 & 0.984& 900 & P & 72\\
2006-01-29   & 01:35 & 0.903& 900 & P & 89\\
\enddata
\tablenotetext{a}{S stands for spectroscopic and P stands for photometric}
\end{deluxetable}

\clearpage

\begin{deluxetable}{lcccccc}
\tablecaption{Magnitudes and colors during eclipses and quadratures}
\tablewidth{0pt}
\rotate
\tablehead{
\colhead{Phase} &\colhead{U} &\colhead{B} &\colhead{V} &\colhead{(R-comp)} &\colhead{U-B} &\colhead{B-V} 
}
\startdata
0.00 (primary eclipse)&  $16.579\pm0.007$ & $15.811\pm0.007$ & $14.479\pm0.06$ & $0.2183\pm0.015$ &
$0.768$ & $1.33$\\
0.25 (Max I) &  $15.739\pm0.007$ & $14.892\pm0.007$ & $13.620\pm0.06$ & $-0.61155\pm0.015$ &
$0.847$ & $1.272$\\
0.50 (secondary eclipse)&  $15.970\pm0.007$ & $15.121\pm0.007$ & $13.829\pm0.06$ & $-0.34813\pm0.015$&
$0.849$ & $1.292$\\
0.75 (Max II) &  $15.595\pm0.007$ & $14.797\pm0.007$ & $13.517\pm0.06$ & $-0.68721\pm0.015$&
$0.798$ & $1.280$\\
\enddata
\end{deluxetable}

\clearpage

\begin{deluxetable}{ccc}
\tablecaption{O'Connell effect}
\tablewidth{0pt}
\tablehead{
\colhead{\ } & \colhead{$MaxII-MaxI$} & \colhead{color}
}
\startdata
$\Delta m_U$ &   -0.144 & \ \\
$\Delta m_B$ &   -0.095 & \  \\
$\Delta m_V$ & -0.103 & \ \\
$\Delta m_R$  & -0.0757& \ \\
$\Delta m_U - \Delta m_B$ & \ & -0.049\\
$\Delta m_B - \Delta m_V$ & \ &  0.008\\
$\Delta m_V - \Delta m_R$ & \ & -0.027\\
\enddata
\end{deluxetable}

\clearpage

\begin{deluxetable}{ccccc}
\tabletypesize{\scriptsize}
\tablecaption{Nightfall Models}
\tablewidth{0pt}
\tablehead{
\colhead{Parameter} & \colhead{Spotless} & \colhead{One Spot} & \colhead{Three spot} 
&\colhead{Uncertainty}
}
\startdata
Period (days)\tablenotemark{a}          & 0.251& 0.251& 0.251& \  \\
i (\degr)                        & 83.3&  83.3 & 83.3 & $\pm$2.3  \\
q (=M$_2$/M$_1$)  & 0.527& 0.527 &  0.527& $\pm$0.030 \\
Total Mass (M$_\sun$)\tablenotemark{a}   & 0.9 & 0.9 & 0.9&$\pm$0.1  \\
T$_1$  (K) \tablenotemark{a}                  & 4100 & 4100 & 4100& $\pm$25 \\
T$_2$  (K)                   & 3170 & 3262 & 3425& $\pm$25\\
Fill Factor$_1$           & 1.00 &1.00  & 1.00& $\pm$ 0.01\\
Fill Factor$_2$           & 0.950  & 0.950 & 0.950& $\pm$0.030\\
Gravity Darkening$_1$ & 0.082 & 0.082 & 0.082& \ \\
Gravity Darkening$_2$ &  -0.016 & -0.002 & 0.020&\ \\
Albedo$_1$ = Albedo$_2$ & 0.5 & 0.5 & 0.5&\ \\
\cutinhead{Spot(s) Primary ({\tt Nightfall} Coordinates)}

Longitude-1 (\degr) & & 272  & 310&\ \\
Latitude-1 (\degr)  & & 0 & 0&\ \\
Radius-1 (\degr) & & 66 & 55& $\pm$10\\
Dimfactor-1 &  &0.985  & 0.992& $\pm$0.007\\
Longitude-2 (\degr) & & & 210&\ \\
Latitude-2 (\degr)& & & 0&\ \\
Radius-2 (\degr) & & & 68& $\pm$10\\
Dimfactor-2 & & & 0.971& $\pm$0.010\\

\tablebreak

\cutinhead{Spot(s) Secondary ({\tt Nightfall} Coordinates)}

Longitude & & & 0&\ \\
Latitude & & & 0&\ \\
Radius (\degr) & & & 45& $\pm$5\\
Dimfactor & & & 0.893& $\pm$0.050\\

\cutinhead{Output Parameters}

${\chi_\nu} ^2$  & 54 & 22 & 9 \\
Mean Temperature$_1$ (K) & 4106 & 4089& 4062&$\pm$25 \\
Mean Temperature$_2$ (K) & 3170 & 3262 & 3365&$\pm$25 \\
Mean Radius$_1$ (R$_\sun$) & 0.673 & 0.673 & 0.673& \\
Mean Radius$_2$ (R$_\sun$) & 0.477 & 0.477& 0.477& \\
Distance (R$_\sun$)  & 1.616 & 1.616& 1.616& \\
Orbital Velocity$_1$ (km/sec) & 111.61 & 111.61& 111.61& \\
Orbital Velocity$_2$ (km/sec) & 211.83 & 211.83 & 211.83& \\
M$_1$ (M$_\sun$) & 0.59 & 0.59 & 0.59 & $\pm$0.02\\
M$_2$ (M$_\sun$) & 0.31 & 0.31 & 0.31& $\pm$0.09 \\
\enddata
\tablenotetext{a}{Held fixed.}
\end{deluxetable}

\begin{deluxetable}{cc}
\tabletypesize{\scriptsize}
\tablecaption{BinaryMaker Model}
\tablewidth{0pt}
\tablehead{
\colhead{Parameter} & \colhead{Three spot}
}
\startdata
Mass ratio & 0.527\\
Fillout 1 & 0.0\\
Fillout 2 & -0.01 \\
Temperature 1& 4100\\
Temperature 2 & 3375\\
G1=G2 & 0.32\\
X1 & 0.887\\
X2 & 0.973 \\
Reflection 1= Reflection 2 & 0.5 \\
Inclination & 81\\
Period & 0.251\\

\cutinhead{Spots Primary ({\tt BinaryMaker} Coordinates)}

Longitude & 330 \\
Latitude & 90 \\
Spot radius & 68.2 \\
Temperature Factor & 0.97\\

Longitude & 230 \\
Latitude & 90\\
Spot radius & 54\\
Temperature Factor & 0.99\\

\cutinhead{Spot Secondary ({\tt BinaryMaker} Coordinates)}

Longitude & 0.0 \\
Latitude & 90 \\
Spot radius & 45 \\
Temperature Factor & 0.89\\

\enddata

\end{deluxetable}

\end{document}